\documentclass[fleqn,usenatbib]{mnras}

\usepackage{newtxtext,newtxmath}

\usepackage[T1]{fontenc}

\DeclareRobustCommand{\VAN}[3]{#2}
\let\VANthebibliography\thebibliography
\def\thebibliography{\DeclareRobustCommand{\VAN}[3]{##3}\VANthebibliography}

\usepackage{graphicx}	
\usepackage{amsmath}	

\newcommand{\schw}{Schwarzschild }
\newcommand{\lam}{$\lambda_{Re, EO}$}
\makeatletter
\newcommand\sendemail[3]{
\edef\@tempa{mailto:#1?subject=#2 }%
\edef\@tempb{\expandafter\html@spaces\@tempa\@empty}%
\href{\@tempb}{#3}}

\catcode\%=11
\def\html@spaces#1 #2{#1
\catcode\%=14
\makeatother

\title[Environment vs orbital structures of SAMI galaxies]{The SAMI Galaxy Survey: Environmental analysis of the orbital structures of passive galaxies}

\author[G. Santucci et al.]{\parbox{\textwidth}{
Giulia Santucci,$^{1,2,3}$\thanks{E-mail: g.santucci@unsw.edu.au}
Sarah Brough,$^{1,3}$
Jesse van de Sande,$^{4,3}$
Richard McDermid, $^{5,3}$
Stefania Barsanti,$^{6,4,3}$
Joss Bland-Hawthorn,$^{4,3}$
Julia J. Bryant,$^{4,7,3}$
Scott M. Croom,$^{4,3}$
Claudia Lagos,$^{2,3}$
Jon S. Lawrence,$^{8}$
Matt S. Owers,$^{5,9,3}$
Glenn van de Ven, $^{10}$
Sam P. Vaughan,$^{5,9,11,3}$
Sukyoung K. Yi$^{12}$
\\}
\vspace{0.4cm}
\\
\parbox{\textwidth}{
$^{1}$School of Physics, University of New South Wales, NSW 2052, Australia\\
$^{2}$ International Centre for Radio Astronomy Research (ICRAR), M468, University of Western Australia, 35 Stirling Hwy, Crawley, WA 6009, Australia\\
$^{3}$ARC Centre of Excellence for All Sky Astrophysics in 3 Dimensions (ASTRO 3D), Australia\\
$^{4}$Sydney Institute for Astronomy, School of Physics, University of Sydney, NSW 2006, Australia\\
$^{5}$Department of Physics and Astronomy, Macquarie University, Sydney, NSW 2109, Australia\\
$^{6}$Research School of Astronomy and Astrophysics, The Australian National University, Canberra, ACT 2611, Australia\\
$^{7}$Australian Astronomical Optics, AAO-USydney, School of Physics, University of Sydney, NSW 2006, Australia\\
$^{8}$Australian Astronomical Optics, Faculty of Science \& Engineering, Macquarie University. 105 Delhi Rd, North Ryde, NSW 2113, Australia\\
$^{9}$Astronomy, Astrophysics and Astrophotonics Research center, Macquarie University, Sydney, NSW 2109, Australia\\
$^{10}$ Department of Astrophysics, University of Vienna, Türkenschanzstrasse 17, 1180 Vienna, Austria\\
$^{11}$Centre for Astrophysics and Supercomputing, School of Science, Swinburne University of Technology, Hawthorn, VIC 3122, Australia.\\
$^{12}$Department of Astronomy and Yonsei University Observatory, Yonsei University, Seoul 03722, Republic of Korea
}
}

\date{Accepted XXX. Received YYY; in original form ZZZ}

\pubyear{2022}

\begin{document}
\label{firstpage}
\pagerange{\pageref{firstpage}--\pageref{lastpage}}
\maketitle

\begin{abstract}
Most dynamical models of galaxies to date assume axisymmetry, which is not representative of a significant fraction of massive galaxies. We have built triaxial orbit-superposition Schwarzschild models of galaxies observed by the SAMI Galaxy Survey, in order to reconstruct their inner orbital structure and mass distribution. The sample consists of 153 passive galaxies with total stellar masses in the range $10^{9.5}$ to $10^{12} M_{\odot}$. We present an analysis of the internal structures and intrinsic properties of these galaxies as a function of their environment. We measure their environment using three proxies: central or satellite designation, halo mass and local $5^{th}$ nearest neighbour galaxy density. We find that although these intrinsic properties correlate most strongly with stellar mass, environment does play a secondary role: at fixed stellar mass, galaxies in the densest regions are more radially anisotropic. In addition, central galaxies, and galaxies in high local densities show lower values of edge-on spin parameter proxy \lam. We also find suggestions of a possible trend of the fractions of orbits with environment for lower-mass galaxies (between $10^{9.5}$ and $10^{11} M_{\odot}$) such that, at fixed stellar mass, galaxies in higher local densities and halo mass have higher fractions of hot orbits and lower fractions of warm orbits. Our results demonstrate that after stellar mass, environment does play a role in shaping present-day passive galaxies.

\end{abstract}




\section{Introduction}
Our current understanding of galaxy formation suggests that massive galaxies form in a two-phase process \citep[e.g.,][]{Naab2009, Oser2010}. During the first phase, at high redshift, they grow by a rapid episode of in-situ star formation, resulting in compact massive systems.
After $z \approx 2$, these massive ($\log_{10} (M_*/ M_{\odot}) > 10.5$) compact galaxies are predicted to be quiescent and grow mostly by accreting mass through galaxy mergers that add stars to their outskirts \citep[e.g.][]{Naab2009, Bezanson2009,Oser2010, vanDokkum2010, Wellons2016}. 

The $\Lambda$CDM model predicts a strong dependence of galaxy properties on their environment \citep[e.g.][]{Springel2005}, in particular because mergers are expected to play a vital role during the formation and/or evolution of almost every massive galaxy \citep[e.g.][] {White1978}. As interactions are more frequent for galaxies in groups compared to isolated galaxies, and since there are a number of physical mechanisms that may act in galaxy clusters to both trigger and truncate star formation in infalling galaxies \citep[for example the interactions between the gas in the galaxy and the hot intra-cluster medium or the gravitational interactions between the galaxy and the cluster's gravitational potential - see][for a review]{Boselli2006, Cortese2021}, a correlation between large-scale environment (groups and clusters, parametrised either as halo mass or position within the halo) and galaxy properties is expected.
In particular, central galaxies are expected to experience additional major mergers, due to their privileged position at the bottom of the potential well of groups \citep[e.g.][]{DeLucia2007}. Similarly, galaxies in dense environments, being exposed to more events that can affect their properties, are expected to show differences compared to galaxies of similar mass in low-density environments. 

The merger history of a galaxy is thought to be one of the major factors that determines its internal kinematic structures \citep[e.g.,][]{White1979, Fall1980, Lagos2018,Park2019,Lagos2022}. Simulations of different combinations of minor and major mergers have been shown to lead to very different types of galaxies \citep[e.g.][]{Naab2014, Lagos2022}. Therefore, we expect the internal orbital structures of central and satellite galaxies to show different characteristics if their merging histories are different. Simulations and observations to date have also shown that central galaxies tend to have more hot orbits and more prolate shapes \citep{Tsatsi2017, Li2018a} than satellite galaxies, consistent with this picture. 

The advent of integral field spectroscopy (IFS) and large IFS surveys, such as ATLAS$^{\rm 3D}$ \citep{Cappellari2011}, the Sydney-AAO Multi-Object Integral-Field Spectrograph (SAMI) Galaxy Survey \citep{Croom2012, Bryant2015}, the Calar Alto Legacy Integral Field Area Survey (CALIFA; \citealt{Sanchez2012}), MASSIVE \citep{Ma2014} and the Mapping Nearby Galaxies at Apache Point Observatory (MaNGA) survey \citep{Bundy2015}, have contributed to significantly expand our understanding of galaxy kinematics and their connection to intrinsic galaxy properties
and their environment \citep[e.g.][]{Cappellari2016}.

IFS surveys have unveiled various correlations between the proxy for the spin parameter, $\lambda_{\rm Re}$, which provides a measurement of how rotationally supported a galaxy is, and galaxy properties for ETGs. For example, $\lambda_{\rm Re}$ is observed to be strongly correlated with stellar mass, so that the fraction of galaxies with low $\lambda_{\rm Re}$ (slow-rotating systems, i.e. galaxies whose kinematics are dominated by random motions) increases with increasing stellar mass \citep{Emsellem2011, vandeSande2017a, Veale2017, Brough2017, Wang2020}. 

However, the results obtained from observations on the importance of the environment in shaping slow-rotating galaxies are contradictory. Central galaxies are generally found to be slow-rotating, with a spin parameter lower than that of other galaxies of the same stellar mass \citep{Brough2011,Emsellem2011, Jimmy2013}. \citet{ DEugenio2013, Houghton2013, Scott2014, Fogarty2014,Cappellari2016, Oliva-Altamirano2017,Brough2017, Rutherford2021} and \citet{vandeSande2021b} show that most slow-rotating galaxies live in high density environments, typical of massive groups or galaxy clusters. However, when galaxies were studied at fixed stellar mass, it was not clear whether this environmental trend held above the strong relationship with stellar mass \citep{Veale2017,Brough2017,Greene2017, Wang2020}. More recently, \citet{vandeSande2021b} found, using inclination-corrected $\lambda_{Re}$ values for a large sample of galaxies with a range of morphologies, that among fast-rotating galaxies at fixed stellar mass, satellite galaxies have the lowest values of $\lambda_{Re}$, with isolated-central galaxies having the highest values and group/cluster centrals lying in-between. Similarly, galaxies in high-density environments have lower values of $\lambda_{Re}$, compared to those in low-density environments, at fixed stellar mass. This evidence points to stellar mass being the main driver of the evolution of the central regions of early-type galaxies from fast- to slow- rotating galaxies, with the environment playing a secondary role.

The observed $\lambda_{Re}$ is directly calculated from observed kinematic maps and can only incorporate line-of-sight kinematics. Building triaxial orbit-superposition \schw models, we can separate dynamically-based derived properties of galaxies, such as intrinsic shape, orbital components, velocity anisotropy and inner mass distribution. These dynamical models can therefore provide more insight into the three-dimensional internal kinematic structures underlying the line-of-sight kinematics. Several different implementations of the Schwarzschild
method, with varying degrees of symmetry, have been described \citep[e.g.][]{Cretton1999,Gebhardt2003, Valluri2004,vandenBosch2008, Vasiliev2015, Vasiliev2020, Neureiter2021}. The Schwarschild method has been used to study the internal stellar structure of gobular clusters \citep{vandeVen2006,Feldmeier2017} and galaxies \citep[e.g.,][]{Thomas2007,Cappellari2007, vandeVen2008, Thomas2014, Feldmeier2017, Poci2019,Jin2020,Santucci2022,Pilawa2022}, to model supermassive black holes \citep{vanderMarel1998, Verolme2002,Gebhardt2003,Valluri2004,Krajnovic2009, Rusli2013,Seth2014,Thomas2016, Thater2017,Thater2019,Mehrgan2019, Liepold2020, Quenneville2021, Quenneville2021b, Thater2021}, and has also been used to identify accreted galactic components \citep[e.g.,][]{Zhu2020,Poci2021,Zhu2022}. \citet{Zhu2018a} separated orbits into four different stellar components: a cold component with near circular orbits (with strong rotation), a hot component with near radial orbits (characterised by random motions), a warm component in-between (characterised by weak rotation) and a counter-rotating component (similar to the warm and cold components).

The number of studies focussed on the connection between the internal structure of galaxies and environment to date is very limited. \citet{Jin2020} analysed the intrinsic properties (orbital components, dark matter content and intrinsic shape) of 144 central and satellite early-type galaxies in the MaNGA survey, based on Schwarzschild models. They found no clear difference between the internal orbital structures of central and satellite galaxies. However, they found that, when considering the local density  environments (indicated by neighbour counts), galaxies in denser regions tended to relatively have higher fractions of hot orbits.

In this paper we analyse triaxial dynamical model fits to SAMI Galaxy Survey data (\citealt{Croom2012,Bryant2015,Owers2017, Croom2021}) as a function of environment. We use properties derived from the \schw models in \citet{Santucci2022} to investigate the extent to which environmental processes affect internal galaxy structures and whether the evolutionary histories of central galaxies are different from those of satellite galaxies. SAMI Galaxy Survey data allow us to study the kinematic properties of a statistical sample of galaxies in a range of environments for the first time. The SAMI Galaxy Survey also allows us to better constrain the \schw models by providing information on the higher kinematic moments.

In Section 2 we describe the sample of galaxies and the data available for this analysis. Section 3 presents our results that are then discussed in Section 4. Our conclusions are given in Section 5.
The SAMI Galaxy Survey adopts a $\Lambda CDM$ cosmology with $\Omega_m=0.3$, $\Omega_{\Lambda} = 0.7$, and $H_0 = 70$ km s$^{-1}$ Mpc$^{-1}$.

\section{Data}
\subsection{SAMI}
The Sydney-AAO Multi-object Integral field spectrograph (SAMI) Galaxy Survey is a large, optical Integral Field Spectroscopy (\citealt{Croom2012,Bryant2015}) survey of low-redshift ($0.04 < z < 0.095$) galaxies covering a broad range in stellar mass,  $8 < \log_{10} (M_{\star}/ M_{\odot}) < 12$, morphology and environment. The sample, with $\approx$ 3000 galaxies, is selected from the Galaxy and Mass Assembly Survey (GAMA survey; \citealt{Driver2011}) equatorial regions (group galaxies), as well as eight additional clusters to probe higher-density environments \citep{Owers2017}.

The SAMI instrument \citep{Croom2012}, on the 3.9m Anglo-Australian telescope, consists of 13 ``hexabundles" \citep{Bland-Hawthorn2011, Bryant2014}, across a 1 degree field of view. Each hexabundle consists of 61 optical fibres that feed into the AAOmaga spectrograph \citep{Sharp2006}. In the typical configuration, 12 hexabundles are used to observe 12 science targets, with the 13th one allocated to a secondary standard star used for calibration. Moreover, SAMI also has 26 individual sky fibres, to enable accurate sky subtraction for all observations without the need to observe separate blank sky frames.

The raw telescope data is reduced into two cubes using the 2dfDR pipeline \footnote{\href{http://www.aao.gov.au/science/software/2dfdr}{http://www.aao.gov.au/science/software/2dfdr}; \citealt{Croom2004, Sharp2010}} together with a purposely written python pipeline for the later stages of reduction \citep{Sharp2015}.

SAMI data consist of 3D data cubes: two spatial dimensions and a third spectral dimension.
The wavelength coverage is from 3750 to 5750 \AA\ in the blue arm, and from 6300 to 7400 \AA\ 
in the red arm, with a spectral resolution of R = 1812 (2.65 \AA\ full-width half maximum; FWHM) and R = 4263 (1.61 \AA ~FWHM), respectively \citep{vandeSande2017a}, so that two data cubes are produced for each galaxy target. 

Each galaxy field was observed in a set of on average seven 30 minute exposures, that are aligned together by fitting the galaxy position within each hexabundle with a two-dimensional Gaussian and by fitting a simple empirical model describing the telescope offset and atmospheric
refraction to the centroids. The exposures are then combined to produce a spectral cube with regular $0.5^{\prime\prime}$ spaxels, with a median seeing of $2.1^{\prime\prime}$.
More details of the Data Release 3 reduction can be found in \citet{Croom2021}\footnote{Reduced data-cubes and stellar kinematic data products for
all galaxies are available on: \href{https://datacentral.org.au}{https://datacentral.org.au}.}.

\subsection{Schwarzschild models}
In \citet{Santucci2022} we presented the orbit-superposition Schwarzschild model fits to a sample of 161 passive SAMI galaxies. Schwarzschild models allow us to model triaxial stellar systems in three steps: firstly we construct a model for the underlying gravitational potential of the galaxy; secondly we calculate a representative library of orbits in these gravitational potentials; and then we find a combination of orbits that can reproduce the observed kinematic maps and luminosity distribution. These steps are fully described in \cite{vandenBosch2008} and \cite{Zhu2018c}. We also corrected the orbital mirroring, as recommended by \cite{Quenneville2021b}, see also \cite{Thater2022}.

The underlying gravitational potential of the galaxy is constructed with a triaxial stellar component and a spherical dark matter halo. The triaxial stellar component mass is calculated from the best-fit two-dimensional Multi-Gaussian Expansion (MGE; \citealt{Emsellem1994, Cappellari2002}) luminosity density \citep{Deugenio2021}. These MGE fits are calculated using $r-$band photometry from re-analysed Sloan Digital Sky Survey (SDSS; \citealt{York2000}) images for GAMA galaxies, reprocessed as described in \citet{Hill2011}, as well as VST/ATLAS (VLT Survey Telescope - ATLAS; \citealt{Shanks2015}) and SDSS DR9 \citep{Ahn2012} observations for cluster galaxies, with VST/ATLAS data reprocessed as described in \citet{Owers2017}. Each best-fit two-dimensional MGE luminosity density is de-projected assuming the orientation in space of the galaxy, described by three viewing angles ($\theta$, $\phi$, $\psi$), to obtain a three-dimensional luminosity density. The space orientation  ($\theta$, $\phi$, $\psi$) can be converted directly to the intrinsic shape, described by the parameters $p$, $q$, $u$. We leave $p$, $q$, $u$ as free parameters to allow intrinsic triaxial shapes to be fitted. By multiplying a constant stellar mass-to-light ratio $M_{\star}/L$ (note that $M_{\star}/L$ is a free parameter in our modelling) by the 3D luminosity, we obtain the intrinsic mass density of stars. 

A spherical Navarro-Frenk-White (NFW; \citealt{Navarro1996}) halo is adopted. The mass, $M_{200}$ (mass enclosed within a radius, $R_{200}$, where the average density is 200 times the critical density), in a NFW dark matter halo is determined by two parameters, left free in our modelling. These are the concentration parameter, $c$, and the fraction of dark matter within $R_{200}$, $f = M_{200}/ M_{\star}$ (where $M_{\star}$ is the total stellar mass). Our Schwarzschild implementation creates initial conditions for the orbits by sampling from the three integrals of motion (energy $E$, second integral $I_2$ and third integral $I_3$). Each set of orbits is sampled across the three integrals with the following number of points: $n_E \times n_{I_2} \times n_{I_3} = 21 \times 10 \times 7$. We tested increased sampling of the second integral $I_2$, for example increasing $n_{I_2}$ from 10 to 18 and to 40, and re-fitting the models. The best-fit values retrieved by this test are consistent, within the 1$\sigma$ confidence level\footnote{We define a confidence level around the minimum value of the models $\chi^2$ and select all the models whose $\chi^2$ is within that confidence level: $\chi^2 - \chi_{min}^2 < \chi_s^2 \times \sqrt{(N_{obs}-N_{par})}$, with $\chi_s^2 = 2$, $N_{obs} = 4N_{kin}$, as we use $V,\ \sigma, \ h_3$ and $h_4$ as model constraints, and $N_{par}$ is the number of free parameters (6 here).}, with the best-fit values retrieved by our regular runs. We therefore use three sets of $21 \times 10 \times 7$ orbits: a typical set of ($E$, $I_2$, $I_3$), a box orbits set (to compensate for the relatively low fraction of box orbits in the inner region present in the typical set) and a counter-rotating set of also ($E$, $- I_2$, $I_3$).

To test whether this number of orbits is enough to reproduce stable results, we select different sets of orbits, increasing $n_E,  n_{I_2}$ and  $n_{I_3}$. We find that, although we see improvements, i.e. lower residuals in the surface brightness maps, when increasing the number of orbits from, for example $21 \times 10 \times 7$ to $31 \times 18 \times 9$, if we increase to $40 \times 18 \times 9$ there is an increase in the systematic residuals. We find that our derived results are stable if $n_E \times n_{I_2} \times n_{I_3}$ is set to $21 \times 10 \times 7$ or more. In addition, we test different orbital radial ranges for the sampling of the energy $E$ and find that our fitted results are stable to the choice of radial range. 

Once the orbit libraries are created, the orbit weights are determined by fitting the set of orbit-superposition models to the projected and de-projected luminosity density (from MGE fits) and the two-dimensional line-of-sight stellar kinematics (derived by \citealt{vandeSande2017a, vandeSande2017b}). The model and the observed values are then divided by the observational error to undertake a $\chi^2$
comparison. The weights are determined by the \citet{vandenBosch2008} implementation, using the \citet{Lawson1974} non-negative least squares (NNLS) implementation. The best-fit model is defined as the model with minimum kinematic $\chi^2$. In order to ensure that we get the best deprojection fit, we force the software to explore other regions of the parameter space (the parameter search easily gets stuck in local minima) by manually setting the parameters in areas that have yet to be explored. This way we are confident that we retrieve the values for the global minimum. More details on the models set up for our sample can be found in \cite{Santucci2022}.

\subsubsection{Derived intrinsic properties}
From these models we derived the internal orbital structure, inner mass distribution, intrinsic shape, velocity anisotropy and edge-on projected spin parameter $\lambda_{Re, EO}$ for each galaxy. In addition, following \citet{Zhu2018a}, we separated orbits into four different components: a cold component with near circular orbits, a hot component with near radial orbits, a warm component in between and a counter-rotating component, according to their orbit circularity $\lambda_z$ as following:
\begin{itemize}
    \item cold orbits with $\lambda_z > 0.8$;
    \item warm orbits with $0.25 < \lambda_z < 0.8$;
    \item hot orbits with $-0.25 < \lambda_z < 0.25$;
    \item counter-rotating orbits with $\lambda_z < -0.25$
\end{itemize}
where  $\lambda_z \equiv J_z/J_{max}(E)$ around the short $z$-axis, normalized by the maximum of a circular orbit with the same binding energy $E$.
We calculated the fraction of orbits in each component within 1$R_e$.
We have derived dynamically-based intrinsic shapes using the triaxial parameter at 1$R_{\rm e}$, $T_{Re} = (1 - p_{Re}^2)/(1 - q_{Re}^2)$, where $p_{Re}$ and $q_{Re}$ are the intermediate-to-long and short-to-long axis ratios. We separate galaxies into four groups according to this parameter: oblate ($T_{Re} \leq 0.1$), mildly triaxial ($0.1 < T_{Re} < 0.3$), triaxial ($0.3 \leq T_{Re} < 0.8$) and prolate ($T_{Re} \geq 0.8$).
We have analysed the velocity anisotropy by defining the luminosity-weighted velocity anisotropy parameter in spherical coordinates, $\beta_r$, following \cite{Binney2008}: 
\begin{equation}
\beta_r = 1 - \frac{\Pi_{tt}}{ \Pi_{rr}},
\end{equation}
with 
\begin{equation}
\Pi_{tt} = \frac{\Pi_{\theta\theta}+\Pi_{\phi\phi}}{2},
\end{equation}
($r,\theta,\phi$) the standard spherical coordinates, and
\begin{equation}\label{eq:D_kk}
\Pi_{kk} =  \int \rho \sigma_k^2 \,d^3x  = \sum_{n=1}^{N} M_n \sigma_{k,n}^2 
\end{equation}
with $\sigma_k$ the velocity dispersion along the direction $k$ at a given location inside the galaxy. The summation defines how we computed this quantity from our Schwarzschild models. $M_n$ is the mass contained in each of the $N$ polar grid cells in the meridional plane of the model, and $\sigma_{k,n}$ is the corresponding mean velocity dispersion along the direction $k$.

We calculate the value of $\beta_r$ within 1$R_{\rm e}$, excluding the inner regions ($r<2^{\prime\prime}$) since this region is affected by atmospheric seeing. $\beta_r > 0$ indicates radial anisotropy, $\beta_r < 0$ indicates tangential anisotropy and $\beta_r = 0$ indicates isotropy. 

Finally, we have reprojected the best-fit models to an edge-on view and measured, within 1$R_{\rm e}$, the proxy for the spin parameter, $\lambda_{Re, EO}$. The measured $\lambda_{Re, EO}$ is consistent with the independently measured and inclination-corrected $\lambda_{Re}$ measured directly from SAMI stellar kinematics by \citet{vandeSande2021b}.

Uncertainties on the measured values are calculated using Monte Carlo realisations, combined with the 1$\sigma$ confidence levels for the fluctuations from the best-fit model (as described in \citealt{Santucci2022}): we use the Monte Carlo realisation to investigate possible biases in this modelling as a result of using SAMI data, which generally has lower S/N and spatial resolution compared to the data used in previous analyses. We find typical uncertainties of $\sim$10-15\%. 

We have explored how these parameters correlate with galaxy stellar mass in \citet{Santucci2022}. In this paper we now analyse their connection to their environment, once the relationships with stellar mass are taken into account.

\subsection{Stellar Mass}
Stellar masses are estimated assuming a \citet{Chabrier2003} initial mass function, from the $g-$ and the $i-$ magnitudes using an empirical proxy developed from GAMA photometry \citep{Taylor2011, Bryant2015}. For cluster galaxies, stellar masses are derived using the same approach \citep{Owers2017}. The $g-$ and $i-$ magnitudes are taken from SDSS images for GAMA galaxies and VST/ATLAS and SDSS DR9 observations for cluster galaxies.

\subsection{Effective Radius}
The effective radius, $R_{\rm e}$, used here is that of the major axis in the $r$-band. The semi-major axis values were taken from MGE profile fits from the $r-$band photometry by \citet{Deugenio2021}. The images used for the MGE fits are square cutouts with $400^{\prime\prime}$ side, centred on the centre of the galaxy, and the MGE fits are calculated using \texttt{MgeFit} \citep{Cappellari2002} and the regularisation feature described in \cite{Scott2009}.

\subsection{Environment measurements}
In order to explore the role of the environment in shaping galaxy properties, we define three proxies for environment: halo mass, central/satellite/isolated and fifth nearest neighbour local environment density. Each of these are described below.

\subsubsection{Halo mass} 
We use the GAMA Galaxy Group Catalogue (G$^{\rm 3}$C; \citealt{Robotham2011}) to define the galaxy groups in the GAMA \citep{Driver2011} regions of the SAMI Galaxy Survey. In this catalogue, galaxies are grouped using an adaptive friends-of-friends (FoF) algorithm, taking advantage of the high spectroscopic completeness of the GAMA survey ($\sim 98.5\%$; \citealt{Liske2015}). For the GAMA groups we use the halo mass, $M_{200}$, the mass contained within $R_{200}$ provided by GAMA. The halo mass was calculated from the group velocity dispersion, using:
\begin{equation}\label{eq:m200} 
\frac{M_{200}}{h^{\rm -1} \ M_{\odot}} = \frac{A}{\frac{G}{M_{\odot}^{\rm -1} \ \rm km^2 \ \rm s^{-2} \ \rm Mpc}} \left ( \frac{\sigma_{\rm FoF}}{\rm km \ \rm s^{-1}}\right)^2  \frac{\rm R_{\rm FoF}}{h^{-1} \ \rm Mpc}
\end{equation}
and then calibrated using a scaling factor, $A$, determined from comparison to haloes from simulated mock catalogues (see \citealt{Robotham2011}). 
$G$ is the gravitational constant, $G = 4.301 \times 10^{-9}\ M_{\odot}^{-1}\rm \ km^2 \ \rm s^{-2} \ \rm Mpc$, $A$ is the scaling factor, $\sigma_{\rm FoF}$ is the velocity dispersion of the group and $R_{\rm FoF}$ is the projected radius of the group.

It is important to note that, following \citet{Owers2017}, we apply a scaling factor of 1.25 to the cluster halo masses, due to calibration, to be consistent with the GAMA halo masses. We also scale the GAMA halo masses (defined using $H_0 = 100$ km s$^{-1}$ Mpc$^{-1}$) to be consistent with the cosmology used here. 
\subsubsection{Central galaxies}
We define the central galaxies as the most massive galaxy within 0.25 $R_{200}$ \citep[e.g.,][]{Oliva-Altamirano2017,Santucci2020}.
To identify the central galaxies for the group sample, we check that the galaxy identified as the ``iterative central" \citep{Robotham2011} is also the most massive galaxy in the group. This is true for 66/67 groups in our sample. One group has a different galaxy selected as the iterative central and the most massive. For this group, we find that the most massive galaxy within a radius of 0.25 $R_{200}$ is the iterative central galaxy and we select it as the central galaxy.

A similar procedure is carried out to select the central galaxy in the clusters, in order to ensure consistency between the samples. We identify the galaxy that sits closest to the centre of the cluster (cluster centroids are taken from \citealt{Owers2017}) as well as the most massive galaxy in the cluster.  For 3 out of 8 clusters, these galaxies are the same. For the other 5 clusters, we find the most massive galaxy within a radius of 0.25 $R_{200}$. For 2 clusters this galaxy is also the central one, whereas for 3 out of 5 clusters (Abell 168, Abell 2399 and Abell 4038) the most massive galaxy within 0.25 $R_{200}$ is not the galaxy closest to the centre. This is consistent with the dynamical state of these clusters as discussed in \citet{Owers2017} and \citet{Brough2017}. We therefore select the most massive galaxy within 0.25 $R_{200}$ as the central galaxy for these clusters.

All other galaxies in each halo are classified as satellite galaxies. Galaxies which do not belong to any halo, or which belong to haloes with only two members, are classified as isolated galaxies.

\subsubsection{Local galaxy density}
We determine the local environment of galaxies using a nearest-neighbour density estimate to probe the underlying density field, with the assumption that galaxies with closer neighbours are also in denser environments \citep[e.g.][]{Muldrew2012}.

We use the fifth nearest-neighbour local surface density measurement, $\Sigma_5$, to quantify the local environment around SAMI galaxies \citep{Brough2013, Brough2017, Croom2021}. $\Sigma_5$ measurements for both GAMA and cluster galaxies are derived using the projected comoving distance to the 5th nearest neighbour ($d_5^2$) with a velocity limit $V_{\rm lim}=1000 \ \rm km \ s^{-1}$, and $M_{\rm lim}=-18.5 \ \rm mag$, so that all neighbors are within a volume-limited density-defining population that has absolute magnitudes $M_{\rm r} < M_{\rm lim} - Q_{\rm z}$ (we assume $Q_{\rm z} = 1.03$, which is defined as the expected evolution of $M_{\rm r}$ as a function of redshift; \citealt{Loveday2015}):
\begin{equation}
    \Sigma_5 =\frac{5}{\pi d_5^2}
\end{equation}
In GAMA, galaxies for which the fifth nearest neighbour is more distant than the nearest survey boundary may have erroneous environment density measurements. None of the galaxies in our sample have measurements of $\Sigma_5$ affected by this problem.

\subsection{Sample Selection}
For this analysis we are particularly interested in whether there is an environmental dependence to the evolution of passive galaxies. We focus on the passive galaxies to determine how their environment affects their internal orbital structures. For this reason, we select the 161 SAMI passive galaxies (selected using the SAMI spectroscopic classification presented in \citealt{Owers2019}) used in the analysis presented in \citet{Santucci2022}. Galaxies with stellar masses below $\log_{10} (M_{\star}/ M_{\odot}) = 9.5$ were excluded, because the incompleteness of the stellar kinematic sample is larger than 50\% of the SAMI galaxy survey sample observed in this mass range, as well as galaxies where $R_{\rm e} < 2^{\prime\prime}$ (due to their spatial size being smaller than the instrumental spatial resolution). The galaxies in the final sample were selected as the optimal compromise between best quality data and reasonable sample size and correspond to galaxies with 85 Voronoi bins within 1$R_{\rm e}$ and $R_{\rm max}> R_{\rm e}$. 

Since the GAMA group sample includes all galaxy associations with two or more members, to ensure the robustness of our group sample we first select all the SAMI observed galaxies that belong to GAMA haloes with a robustly estimated velocity dispersion and therefore halo mass (i.e. $\sigma > \sigma_{err}$; \citealt{Robotham2011}). These cuts leave a sample of 146 galaxies. Of these, 37 are central and 90 are satellite galaxies and 19 are isolated galaxies (i.e. not assigned to any group). 

In order to further probe the cluster environment, we select an additional 7 central cluster galaxies, whose kinematic measurements did not reach 1 $R_{\rm e} $. Since the measurements for these galaxies did not meet our radial coverage criterion (outlined in \citealt{vandeSande2017a, Santucci2022}), larger uncertainties have been applied to their derived intrinsic properties from the Schwarzschild modelling (see Appendix A in \citealt{Santucci2022}).

\begin{figure*}
\centering
\includegraphics[scale=0.43,trim=4cm 0cm 2.5cm 2cm , clip=True]{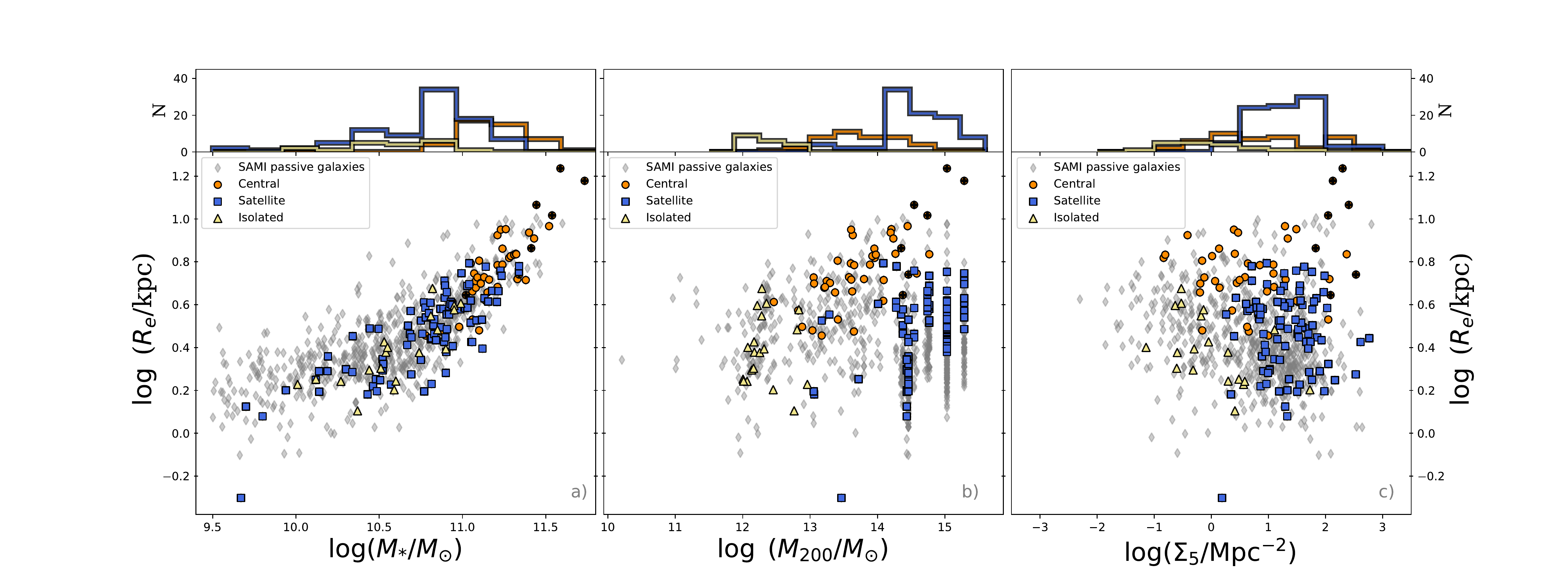}
\caption[Sample distribution in effective radius, $R_e$, versus stellar mass, halo mass and local density]{Effective radius, $R_e$, versus stellar mass (panel a), halo mass (panel b) and local density (panel c). Grey diamonds are the passive galaxies in the SAMI sample with $\log_{10} (M_{\star}/ M_{\odot}) > 9.5$ and $R_{\rm e} > 2^{\prime\prime}$ (738), orange circles are central galaxies (44), blue squares are satellite galaxies (90) and yellow triangles are isolated galaxies (19) in our SAMI sample. The additional 7 central cluster galaxies, whose kinematic measurements did not meet our radial coverage criterion, are highlighted with black crosses. Top panels show the distribution of central/satellite/isolated galaxies in stellar mass, halo mass and local density. }
\label{fig:sample_cen}
\end{figure*}

This inclusion gives us a final sample of 153 SAMI galaxies, including 44 central galaxies, 90 satellite galaxies and 19 isolated galaxies. These are shown in Fig. \ref{fig:sample_cen} and used hereafter in this analysis.

\section{Intrinsic properties of galaxies as function of environment}
\begin{figure}
\centering
\includegraphics[scale=0.28,trim=0.25cm 0.5cm 0.25cm 0cm , clip=True]{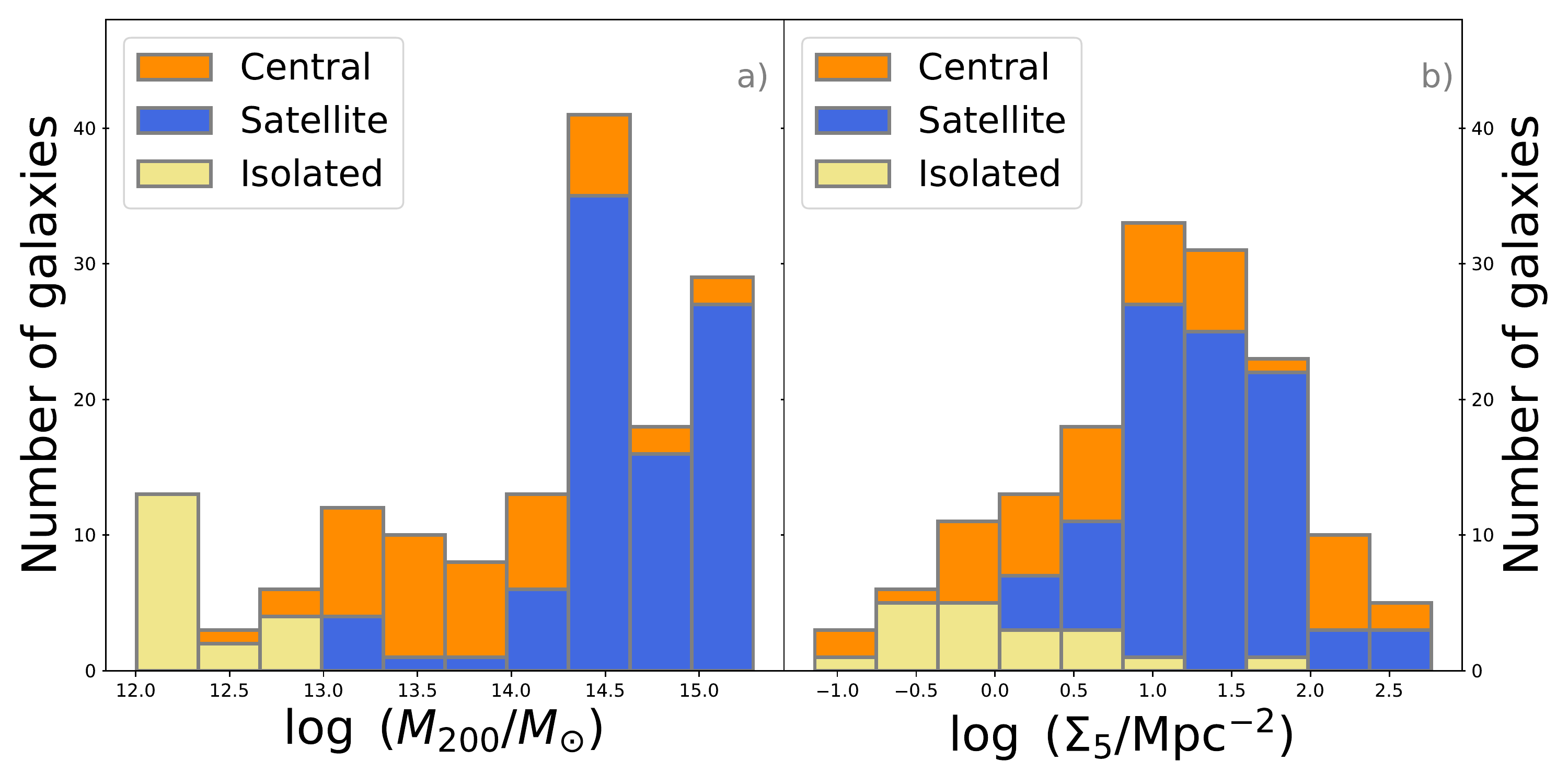}
\caption{Distribution of halo mass ($\log \ M_{200}/M_{\odot}$, panel a) and local density ($\log \Sigma_5$, panel b) for the galaxies in our sample. The cumulative histograms are colour-coded by galaxy designation (central, satellite or isolated).}
\label{fig:distr}
\end{figure}

To explore the role of environment in shaping the intrinsic properties of galaxies, we first divide the sample as follows:
central (44 galaxies), satellite (90 galaxies) and isolated (19 galaxies). Each of these have been divided into four mass bins, with an equal number of galaxies in each (11 for centrals, 22 for satellites and 5 for isolated galaxies) In Sec. \ref{cen_sat} we show how intrinsic properties such as edge-on projected \lam, triaxiality, velocity anisotropy and fraction of orbits change with galaxy designation. We then explore how these properties and the fractions of orbits are distributed in the halo mass-stellar mass and local density-stellar mass planes in Sec. \ref{halo_density}. We show the distribution of the galaxies in our sample with halo mass (panel a) and local density (panel b) in Fig. \ref{fig:distr}. 

\subsection{Intrinsic properties of central, satellite and isolated galaxies}\label{cen_sat}
\begin{figure*}
\centering
\includegraphics[scale=0.44,trim= 4.1cm 0.3cm 4cm 1cm , clip=True]{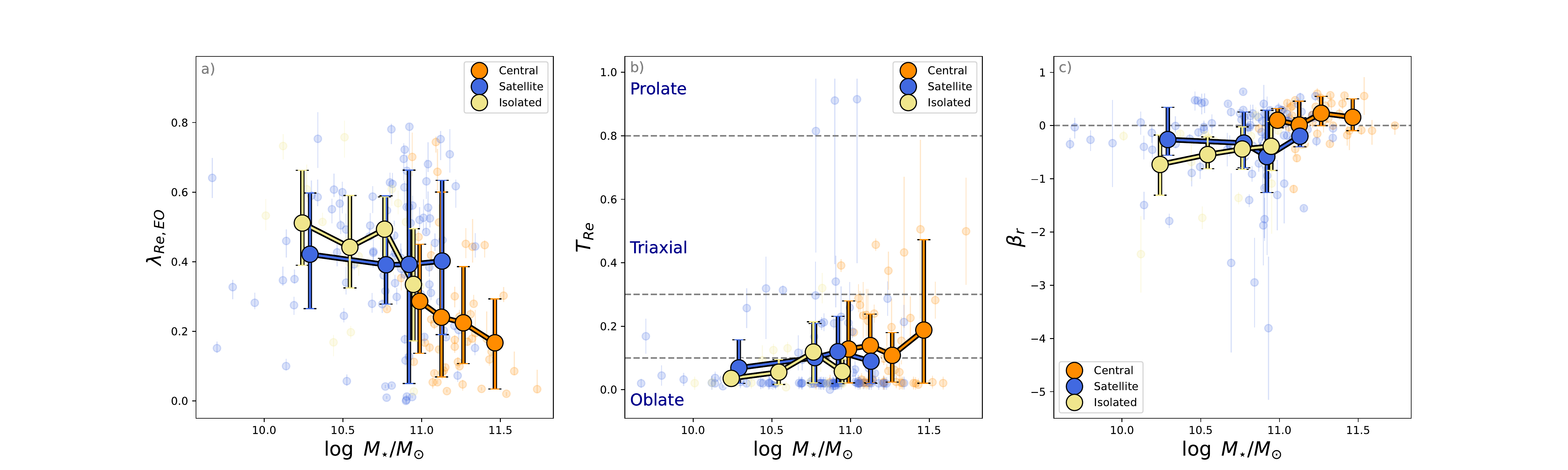}
\caption{Spin parameter $\lambda_{Re, EO}$ (panel a), , triaxiality $T_{Re}$ (panel b) and velocity anisotropy $\beta_r$ (panel c) as a function of stellar mass. The mean values of $\lambda_{Re, EO}$, $T_{Re}$ and $\beta_r$ for 4 mass bins are shown as large circles, colour-coded by central/satellite/isolated designation. Bold error bars represent the 1$\sigma$ scatter around the mean value of each mass bin. At fixed stellar mass, central galaxies have consistently lower mean values of $\lambda_{Re, EO}$ than satellite galaxies, and show a strong trend with stellar mass. Central galaxies are more likely to be triaxial than satellite and isolated galaxies and they are dominated by radial anisotropy. Satellite galaxies do not show any trend of $\lambda_{Re, EO}$, or $T_{Re}$, with mass. Satellite and isolated galaxies are dominated by tangential anisotropy. }
\label{fig:lambda_mass}
\end{figure*}

We present the spin parameter measured from the reprojected edge-on maps $\lambda_{Re, EO}$ as a function of stellar mass in Fig. \ref{fig:lambda_mass}, panel a. Dividing the galaxies into central, satellite and isolated galaxies, we find suggestions that at fixed stellar mass (for galaxies with $\log M_{\star}/M_{\odot} \geq 11$), central galaxies have consistently lower mean values of $\lambda_{Re, EO}$ than satellite galaxies, which do not show any trend of $\lambda_{Re, EO}$ with stellar mass. We note, however, that the mass range where these satellite and central galaxies overlap is small. Isolated and central galaxies are more slowly rotating with increasing stellar mass. 

We then explore galaxy triaxiality as a function of stellar mass for central, satellite and isolated galaxies in Fig. \ref{fig:lambda_mass}, panel b. Central galaxies are more likely to be triaxial (14\% $\pm$ 5\% of the central galaxies have $T_{Re} > 0.3$), compared to satellite and isolated galaxies (7\% $\pm$ 3\% and 5\% $\pm$ 4\%, respectively). However, this increase in the fractions of triaxial shapes for central galaxies is likely driven by the stellar mass as more massive galaxies have higher mean triaxiality (as also seen in \citealt{Jin2020}, Fig. 11, and \citealt{Santucci2022}, Fig. 8). We also note that the three most prolate-like galaxies in our sample are satellite galaxies.

In Fig. \ref{fig:lambda_mass}, panel c, we show the velocity anisotropy parameter in spherical coordinates, $\beta_r$, calculated within 1$R_e$ as described in \citet{Santucci2022}, as a function of stellar mass for central/satellite/isolated galaxies. $\beta_r > 0$ indicates that the galaxy is dominated by radial anisotropy, $\beta_r < 0$ indicates tangential anisotropy and $\beta_r = 0$ indicates isotropy. We find that, at fixed stellar mass, central galaxies are generally more radially anisotropic than satellite galaxies and that isolated galaxies show lower values of $\beta_r$ (more tangentially anisotropic) than satellite galaxies, but no difference is found at stellar masses greater than $\log \ M_{\star}/M_{\odot} \sim 10.75$.

\begin{figure*}
\centering
\includegraphics[scale=0.5,trim= 1.5cm 2cm 1cm 1cm , clip=True]{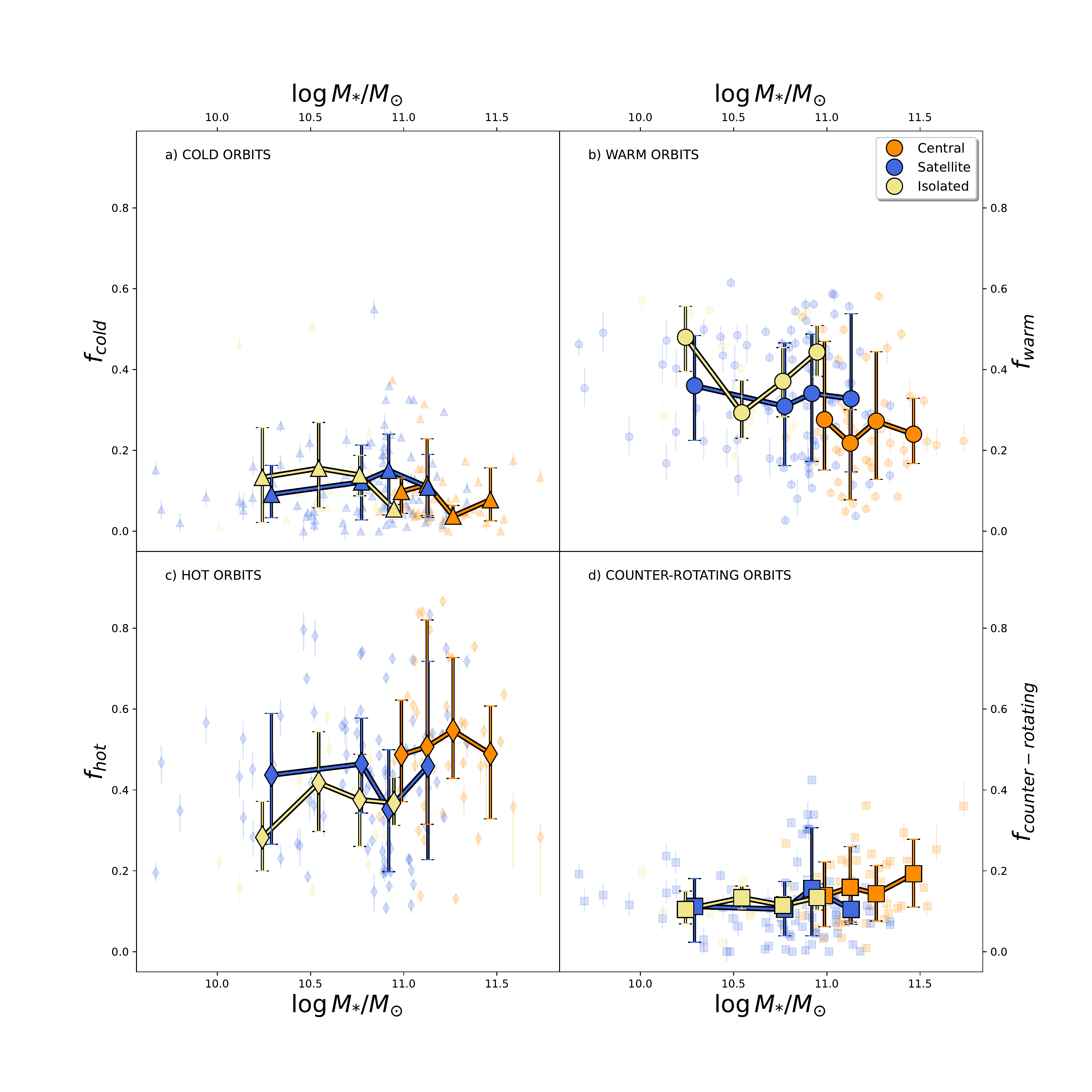}
\caption[Fractions of: a) cold orbits, b) warm orbits, c) hot orbits and d) counter-rotating orbits as a function of stellar mass for different environments]{Fractions of: a) cold orbits, b) warm orbits, c) hot orbits and d) counter-rotating orbits as a function of stellar mass. Galaxies are classified into central (orange), satellite (blue) and isolated (yellow) galaxies. Each class is divided into 4 mass bins, with the bold points representing the mean values for each mass bin and the error bars indicate the 1$\sigma$ scatter around the mean value. In general, central, satellite and isolated galaxies show similar trends for the fraction of orbits, with no significant difference, although at fixed stellar mass, central galaxies have the lowest fractions of warm orbits and the highest fractions of hot orbits.}
\label{fig:orbs_cen}
\end{figure*}

In \citet{Santucci2022}, we divided each galaxy into four orbital components, according to their orbit circularity distribution (cold, warm, hot and counter-rotating components), calculating the fraction of each component within 1$R_e$. We show the four orbital fractions as a function of stellar mass divided into central, satellite and isolated galaxies in Fig. \ref{fig:orbs_cen}. 
Central galaxies have the lowest fractions of warm orbits and the largest fractions of hot orbits at all masses. At fixed stellar mass, central galaxies have slightly lower mean fractions of warm orbits, compared to satellite and isolated galaxies of similar mass. No other significant difference is found between the orbital fractions of central, satellite and isolated galaxies.

\subsection{Environment-stellar mass plane} \label{halo_density}
To better visualise the potential trends with environment, we now explore the distribution of the edge-on projected $\lambda_{Re, EO}$, triaxiality, velocity anisotropy and orbital fractions in the halo mass-stellar mass and local density-stellar mass planes, in order to constrain the role of the environment once the relationship between these parameters and stellar mass is taken into account.

\subsubsection{Spin parameter}
\begin{figure*}
\centering
\includegraphics[scale=0.5,trim= 2cm 3cm 1cm 3cm , clip=True]{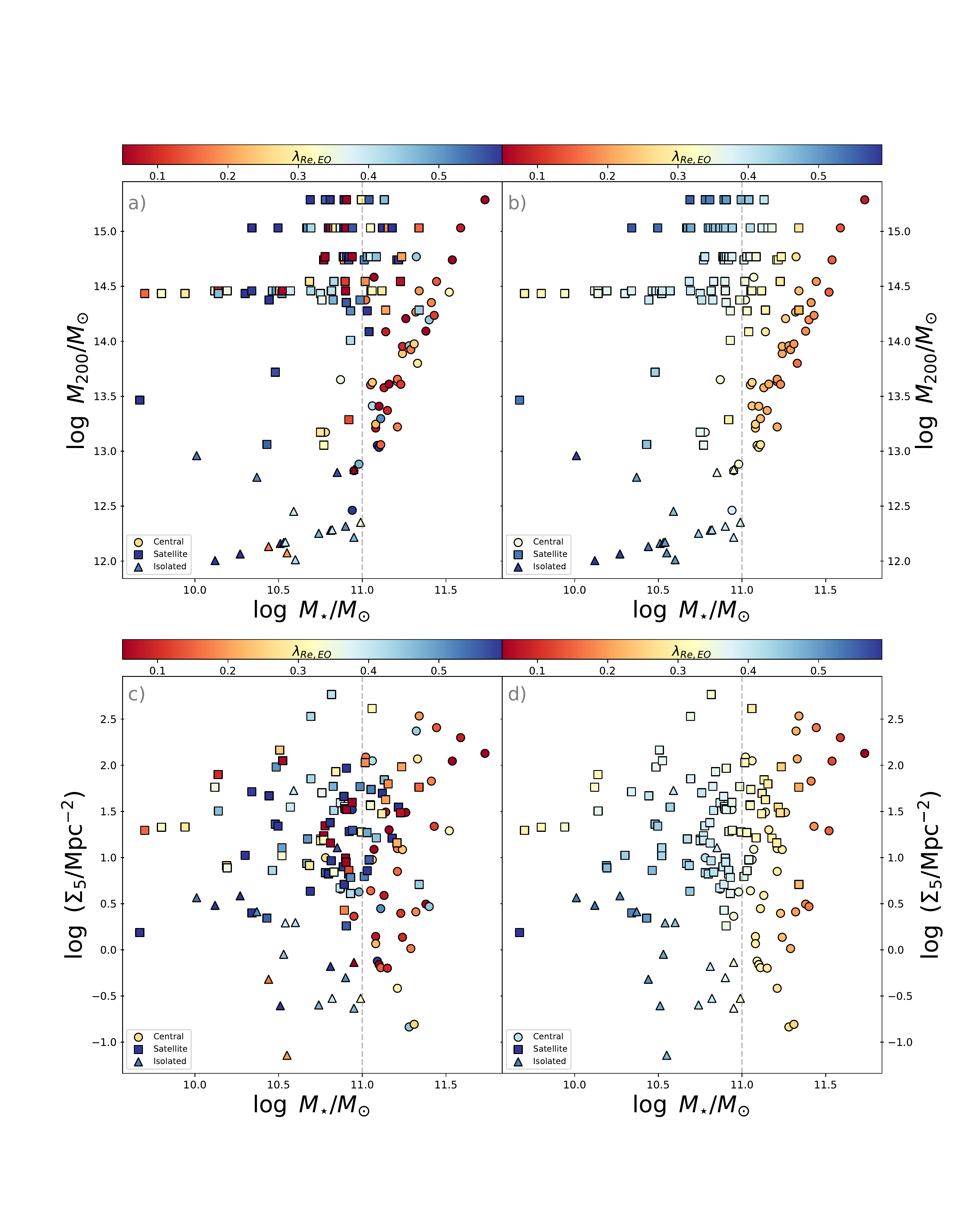}
\caption[Halo mass $\log \ M_{200}/M_{\star}$ and local density $\log \Sigma_5$ as a function of stellar mass, colour-coded by the intrinsic spin parameter \lam]{Halo mass $\log \ M_{200}$ and local density $\log (\Sigma_5/ \rm Mpc^{-2})$ as a function of stellar mass, colour-coded by the intrinsic spin parameter \lam (left-hand panels a, c), and LOESS smoothed (right-hand panels b, d). We find that at fixed environment, $\lambda_{Re, EO}$ decreases with increasing stellar mass. At fixed stellar mass, for masses above $\log M_{\star}/M_{\odot} \sim 11$, there is weak evidence of a relationship between $\lambda_{Re, EO}$ and halo mass. $\lambda_{Re, EO}$ also decreases with increasing local density.}
\label{fig:lam_r_env}
\end{figure*}

We explore the environmental dependence of the spin parameter measured from the reprojected edge-on maps $\lambda_{Re, EO}$ as a function of stellar mass and environment in Fig. \ref{fig:lam_r_env}, colour-coding the galaxies by their edge-on spin parameter \lam (panel a and c). We apply a locally weighted regression algorithm (LOESS - \citealt{Cappellari2013a}) to the data to recover any mean underlying trend (shown in panel b and d). In order to quantify the trends with environment, we calculate, from the native data, the Spearman's semi-partial correlation coefficient, $\rho$, using the Python package \textit{pinguoin.partial\_corr} \citep{Vallat2018}. This allows us to calculate the strength of a correlation of our parameters ($\lambda_{Re, EO}$, in this case) with the environment, taking into account the contribution from stellar mass, and to determine how likely it is that any observed correlation is due to chance. A $\rho$ value close to 1 indicates a strong correlation, while a value close to $-$1 indicates a strong anti-correlation \footnote{We present all relationships with probabilities 1-$\sigma$ or greater, as these analyses are among the firsts of their kind, although the sample is small for bimodal statistics. We note that even though the relationships do not show strong correlations, they are important to guide future analyses.}. At fixed stellar mass, for masses above $\log M_{\star}/M_{\odot} \sim 11$, there is a suggestion of a relationship between $\lambda_{Re, EO}$ and halo mass so that galaxies in lower-mass haloes are less rotationally supported (Spearman's semi-partial correlation coefficient $\rho$ = - 0.22, significant at the 1$\sigma$ level, with a $p$-value of 0.084). This region of the stellar mass - halo mass plane is dominated by central galaxies, which are, at fixed stellar mass, less rotationally supported (as seen in Fig. \ref{fig:lambda_mass}), suggesting that this trend is likely be driven by the central/satellite/isolated designation. For galaxies with stellar masses below $\log M_{\star}/M_{\odot} \sim 11$ we do not find any significant correlation with halo mass. When looking at the trends with local density we find that, at stellar mass $\log M_{\star}/M_{\odot} < 11$, $\lambda_{Re, EO}$ decreases with increasing local density ($\rho$ = -0.27, significant at the 1$\sigma$ level, with a $p$-value of 0.097), so that we see suggestions that galaxies in higher local densities are less rotationally supported than galaxies in less dense regions. 

\begin{figure*}
\centering
\includegraphics[scale=0.15,trim= 3cm 7cm 1cm 7cm , clip=True]{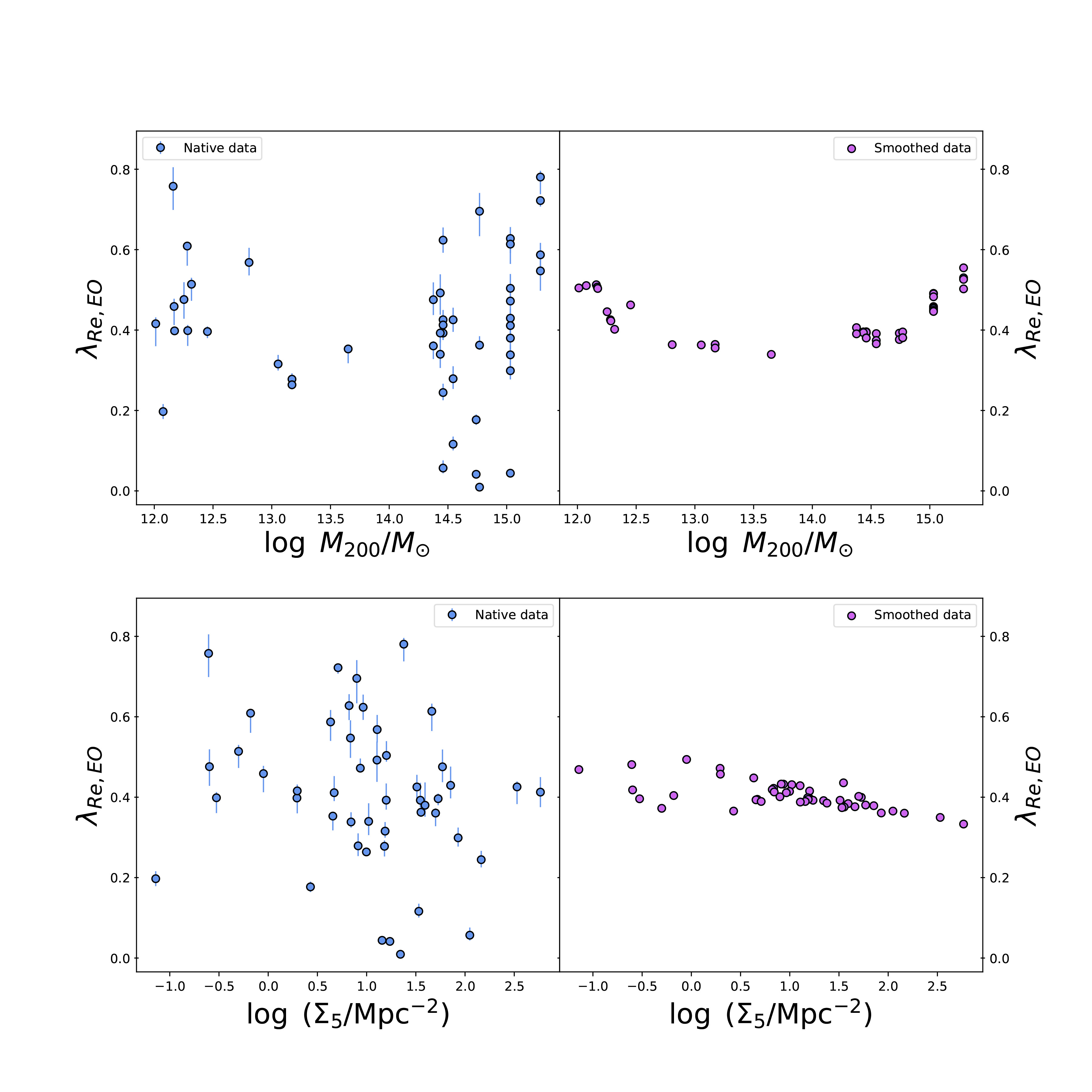}
\caption{Intrinsic spin parameter $\lambda_{Re, EO}$ as a function of halo mass $\log \ M_{200}$ (top rows) and local density $\log (\Sigma_5/ \rm Mpc^{-2})$ (bottom rows). left-hand panels show the native data (in blue), while the LOESS smoothed data (in purple) is shown in the right-hand panels. The smoothed data is representative, within the uncertainties, of the native data.}
\label{fig:lam_1dcut}
\end{figure*}

To better visualise how well the smoothed data reproduce the native data, within the given uncertainties, we show in Fig. \ref{fig:lam_1dcut} a one-dimensional cut of $\lambda_{Re, EO}$ as a function of halo mass (top row) and as a function of local density (bottom row) for galaxies with stellar mass $\log M_{\star}/M_{\odot} = 10.7 \pm 0.25$. The left-hand panels show the native data (in blue), while the right-hand panels show the smoothed values (in purple) of $\lambda_{Re, EO}$. The smoothed data reproduces the general trend of the native data, although with less scatter.

\subsubsection{Intrinsic shape}
We show the distribution of galaxy triaxiality in the halo mass-stellar mass and local density-stellar mass plane in Appendix \ref{app:T_re}, Fig. \ref{fig:app_triax_env}. 
In both cases, we do not find any significant correlation between triaxiality and environment.

\subsubsection{Velocity anisotropy}
\begin{figure*}
\centering
\includegraphics[scale=0.5,trim=2cm 3cm 1cm 3cm , clip=True]{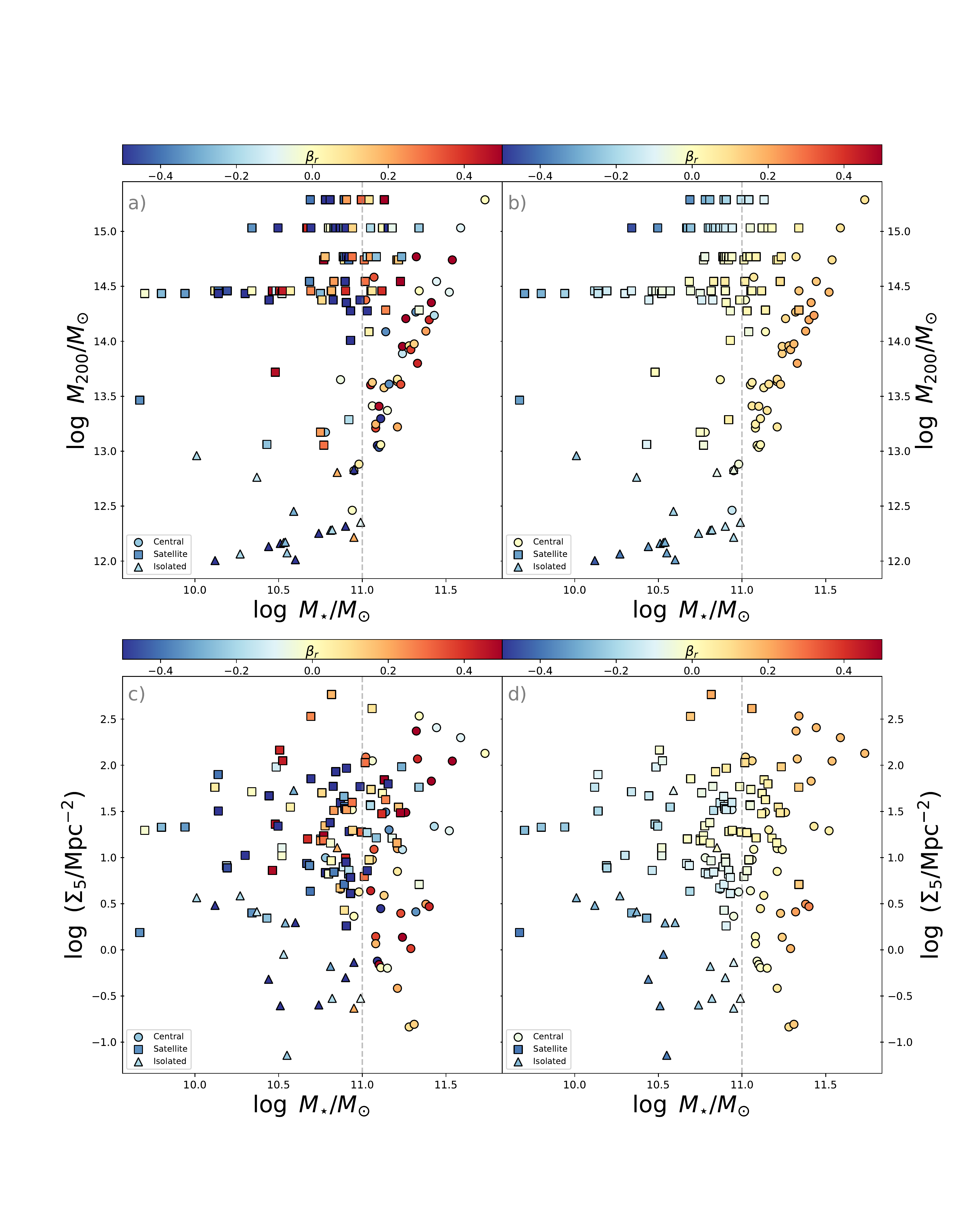} 
\caption{Halo mass $\log \ M_{200}/M_{\odot}$ and local density $\log (\Sigma_5/ \rm Mpc^{-2})$ as a function of stellar mass, colour-coded by the velocity dispersion anisotropy in spherical coordinates $\beta_r$ (panel a and c), and LOESS smoothed (panel b and d) to reveal any mean underlying trend. Central galaxies are shown as circles, satellite galaxies as squares and isolated galaxies as triangles. We find that at fixed halo mass (panel a and b), $\beta_r$ increases with increasing stellar mass. Similarly, at fixed stellar mass, $\beta_r$ increases with increasing halo mass. However, the most radially anisotropic galaxies are in mid-mass haloes. Looking at the local density (panel c and d), we find that at fixed local density, $\beta_r$ increases with increasing stellar mass. Similarly, at fixed stellar mass, $\beta_r$ increases with increasing local density.}
\label{fig:beta_r_env}
\end{figure*}

We explore the distribution of our galaxies in halo mass in Fig. \ref{fig:beta_r_env} (panel a) and local density (panel c) as a function of stellar mass, colour-coding the galaxies by their velocity anisotropy and LOESS smoothed (panel b and panel d) to recover any mean underlying trend.

At fixed environmental proxy, $\beta_r$ increases with increasing stellar mass. We find some hints that the most radially anisotropic galaxies are in mid-mass haloes. This could be a selection effect, since the most radially anisotropic galaxies are central galaxies and we only have 8 central galaxies in clusters (out of 44 central galaxies in the sample, e.g. Fig. \ref{fig:lambda_mass}, panel c). For galaxies with stellar masses below $\log \ M_{\star}/M_{\odot} \sim 11$, we find a value of $\rho = 0.19$ for a correlation between $\beta_r$ and halo mass taking stellar mass into account. This is significant at the 1-$\sigma$ level (with a $p$-value of 0.767). For local density (panel c and d), we find evidence of a correlation of $\beta_r$ with environment so that at fixed stellar mass galaxies in higher local densities have greater values of $\beta_r$, particularly visible for galaxies with stellar masses $\log \ M_{\star}/M_{\odot} < 11$. We find $\rho = 0.33$ for a correlation between $\beta_r$ and local density - significant at the 2-$\sigma$ level (with a $p$-value o 0.035). This suggests the possibility of a weak positive correlation between velocity anisotropy and halo mass and local density such that velocity dispersion anisotropy increases with increasing environmental density. 

A one-dimensional cut of $\beta_r$ as a function of halo mass and as a function of local density for galaxies with stellar mass $\log M_{\star}/M_{\odot} = 10.7 \pm 0.25$ is shown in Fig. \ref{fig:br_1dcut}. 

\subsubsection{Orbital components}
\begin{figure*}
\centering
\includegraphics[scale=0.5,trim=2cm 3cm 1cm 3cm , clip=True]{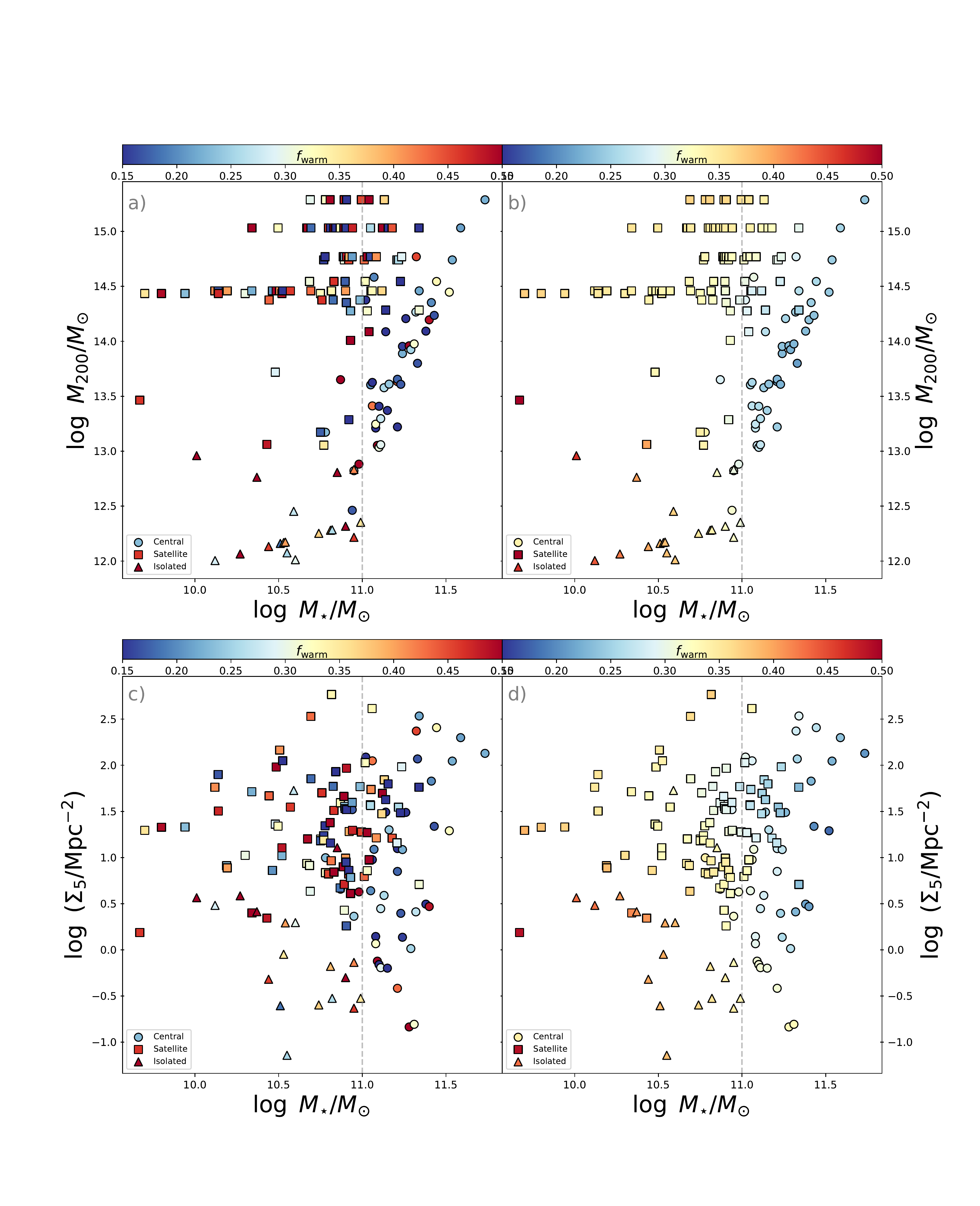} 
\caption{Halo mass $\log \ M_{200}/M_{\odot}$ and local density $\log (\Sigma_5/ \rm Mpc^{-2})$ as a function of stellar mass, colour-coded by the fraction of warm orbits (panel a and c), LOESS smoothed (panel b and d) to reveal any mean underlying trend. Central galaxies are shown as circles, satellite galaxies as squares and isolated galaxies as triangles. We find suggestions that at fixed stellar mass, galaxies with stellar mass below $\log \ M_{\star}/M_{\odot} \sim 11$ have lower fractions of warm orbits with increasing halo mass (panel a and b) and increasing local density (panel c and d).}
\label{fig:f_warm_env}
\end{figure*}

\begin{figure*}
\centering
\includegraphics[scale=0.5,trim= 2cm 3cm 1cm 3cm, clip=True]{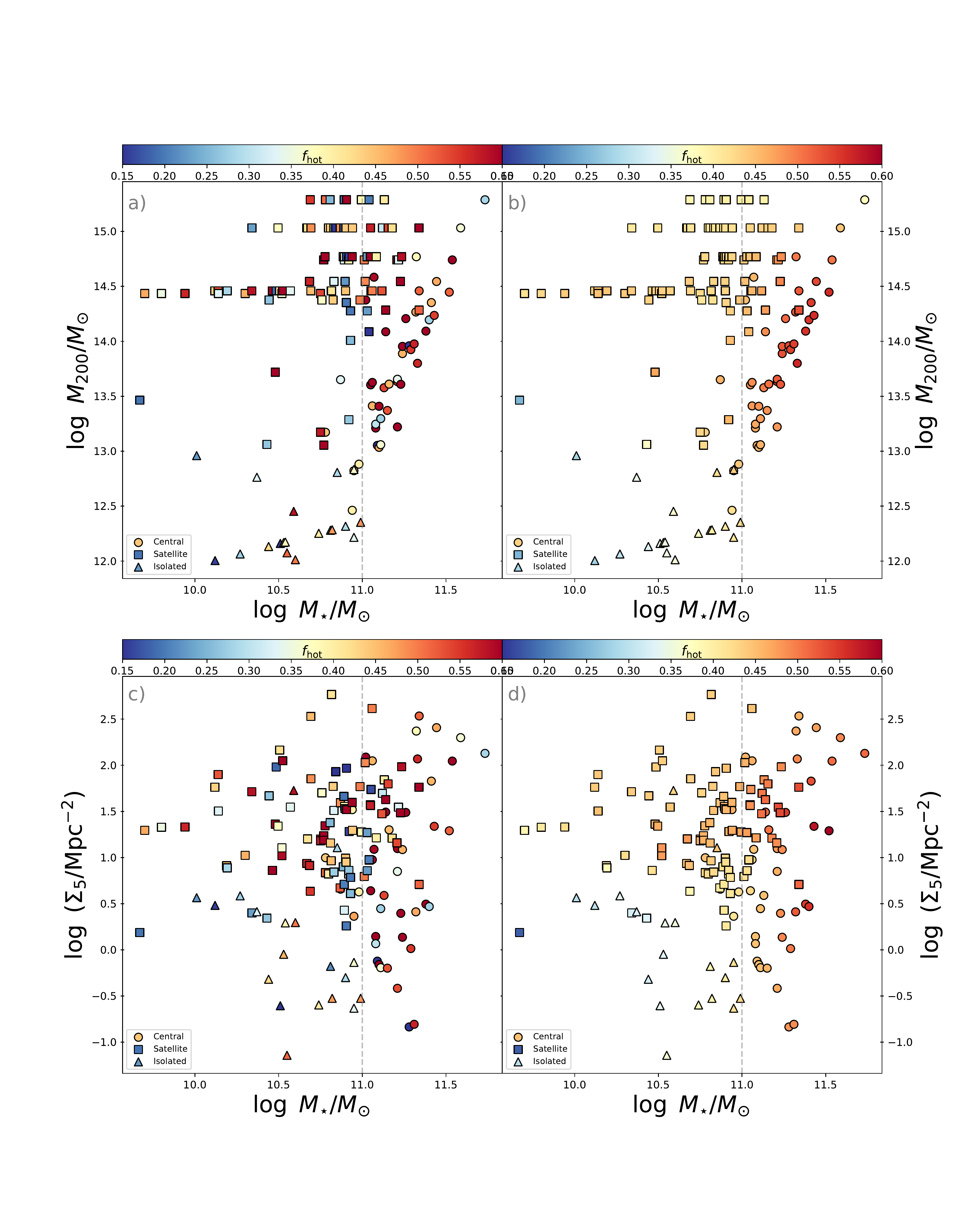} 
\caption{Halo mass $\log \ M_{200}/M_{\odot}$ and local density $\log (\Sigma_5/ \rm Mpc^{-2})$ as a function of stellar mass, colour-coded by the fraction of hot orbits (panel a and c), LOESS smoothed (panel b and d) to reveal any mean underlying trend. Central galaxies are shown as circles, satellite galaxies as squares and isolated galaxies as triangles. We find suggestions that at fixed stellar mass, galaxies with stellar mass below $\log \ M_{\star}/M_{\odot} \sim 11$ have higher fractions of hot orbits with increasing halo mass (panel a and b) and increasing local density (panel c and d).}
\label{fig:f_hot_env}
\end{figure*}

We now explore the distribution of the fraction of orbits in the halo mass-stellar mass and local density-stellar mass planes. We show halo mass (panel a and b) and local density (panel c and d) as a function of stellar mass, colour-coded by the fraction of warm orbits in Fig. \ref{fig:f_warm_env} and hot orbits in Fig. \ref{fig:f_hot_env}.  

We find suggestions of a possible relationship between the fraction of warm orbits and environment (Fig. \ref{fig:f_warm_env}), so that, at fixed stellar mass, the fraction of warm orbits for galaxies with stellar mass below $\log M_{\star}/M_{\odot} \sim 11$ decreases with increasing halo mass and increasing local density. The Spearman's semi-partial correlation coefficient gives a suggestion of a trend ($\rho = -0.19$ at the 1-$\sigma$ level, with a $p$-value of 0.077) for the fraction of warm orbits with local density.

For stellar masses below $\log M_{star}/M_{\odot} \sim 11$, galaxies show evidence of a possible additional relationship between the fraction of hot orbits and environment so that galaxies in higher-mass haloes and denser local environments are more likely to have higher fractions of hot orbits (Fig. \ref{fig:f_hot_env}). Testing the correlation of the fractions of hot orbits for galaxies with stellar mass below $\log M_{star}/M_{\odot} \sim 11$ we find a Spearman's semi-partial correlation coefficient $\rho = 0.22$ at the 1-$\sigma$ level, with a $p$-value of 0.18 with halo mass, and $\rho$ = 0.18 at the 1-$\sigma$ level, with a $p$-value of 0.089, when considering the local density. 

A one-dimensional cut of the fractions of hot and warm orbits as a function of halo mass and as a function of local density for galaxies with stellar mass $\log M_{\star}/M_{\odot} = 10.7 \pm 0.25$ is shown in Fig. \ref{fig:fhot_1dcut} and \ref{fig:fwarm_1dcut}.

We note that there is significant scatter in the distribution of the fraction of orbits, at fixed mass, with environment, similar to what we see in the distribution of the orbits with stellar mass (e.g., Fig. \ref{fig:orbs_cen}). This scatter could be caused by an additional underlying distribution, driven by other parameters. A larger sample is needed to further explore what causes the large scatter in the distribution of the fraction of orbits with environment.

We do not find any additional correlation between the fraction of cold/counter-rotating orbits and halo mass or between the fraction of cold/counter-rotating orbits and local density. We show the distributions of the fraction of cold orbits in Fig. \ref{fig:f_cold_env} and of the fraction of counter-rotating orbits in Fig. \ref{fig:f_cc_env}, in Appendix \ref{app:frac_coldcc} for completeness.
\section{Discussion}

We have explored the connection between environment and galaxy dynamically-derived properties such as intrinsic shape, velocity anisotropy and orbital components using three proxies for environment: central/satellite/isolated, halo mass and fifth nearest neighbour local environment density. We analysed a sample of 153 passive galaxies from the SAMI Galaxy Survey, which allows us to study dynamically-derived galaxy properties for a significant number of galaxies in a range of environments for the first time. Here we discuss how our results compare to the literature and possible evolution scenarios for passive galaxies..

\subsection{Spin Parameter}
Exploring the connection between the intrinsic edge-on $\lambda_{Re}$ and the environment we found that galaxies in high local densities showed lower values of $\lambda_{Re}$, at fixed stellar mass, for galaxies with $\log M_{\star}/M_{\odot} < 11$ (Fig. \ref{fig:lam_r_env}). Moreover, central galaxies (with stellar mass $\log M_{\star}/M_{\odot} > 11$) have lower values of $\lambda_{Re}$ than satellite galaxies of similar stellar mass (Fig. \ref{fig:lambda_mass}). 

Previous results in the literature have found a strong dependence of $\lambda_{Re}$ with stellar mass. However, there are contradictory results for whether the environment plays a role in shaping galaxies, in particular slow-rotating ones. Several studies found a higher fraction of slow-rotating galaxies in the densest environments \citep[e.g][]{Cappellari2011b, DEugenio2013, Scott2014, Fogarty2014}. However, later studies controlled for stellar mass and found that the environment has no significant correlation with the fraction of slow-rotating galaxies \citep{Brough2017, Veale2017, Green2018}. These differences were resolved in \citet{vandeSande2021b} and explained by sample sizes and differences in selection and range in environment studied.

Our results are in agreement with \citet{vandeSande2021b}, who derived the intrinsic values of $\lambda_{Re}$ by applying corrections for seeing, aperture effects and inclination for $\sim$ 1800 SAMI galaxies, including ETGs and LTGs, following \citet{Harborne2020}. They found that central galaxies at stellar masses greater than $\log M_{\star}/M_{\odot} \sim 11$ are more likely to be slow-rotators than satellite galaxies in the same mass bin. Moreover, they also find a similar trend of $\lambda_{Re}$ with local density as we observe here.

There is also evidence from simulations that points towards environmental dependence as a weaker secondary effect on $\lambda_{Re}$. For example,
\citet{Lagos2018}, using IFS-like “observations” of synthetic galaxies from the EAGLE and HYDRANGEA cosmological hydrodynamic simulations, confirmed the primary dependence of $\lambda_{Re}$ on stellar mass, but find a weak, secondary dependence on environment so that, at fixed stellar mass ($\log M_{\star}/M_{\odot} \sim 11.25$), central galaxies have lower values of $\lambda_{Re,EO}$, compared to satellite galaxies, in agreement with our results. \citet{Choi2018}, with Horizon-AGN simulation data, found a weak trend between the spin parameter and environment, so that galaxies in denser environments (higher-mass haloes) rotate more slowly. They note that the trend is driven by satellite galaxies. Although we do not see a trend with halo mass, we do see a similar trend with local density. This difference is likely explained by the fact that most of our satellite galaxies are in cluster environments, while there are fewer group satellite galaxies present in our sample for a clear comparison. In their simulations, \citet{Choi2018} found that galaxy mergers (both major and minor) appear to be the main cause of the spin changes for the majority of the central galaxies, while satellite galaxies are more likely to undergo a spin-down from non-merger tidal perturbation (due to high-speed encounters). 

\subsection{Intrinsic shape}
We find that, in general, central galaxies are more likely to be triaxial ($T_{Re} > 0.3$) than satellite and isolated galaxies: 14\% $\pm$ 5\% of central galaxies compared to 7\% $\pm$ 3\% and 5\% $\pm$ 4\% of satellite and isolated galaxies, respectively. However, at fixed stellar mass, there is no significant difference in the mean triaxiality of central, satellite and isolated galaxies (Fig. \ref{fig:lambda_mass}, panel b). This is consistent with previous observations by \citet{Jin2020}, who do not find any significant differences between the intrinsic shapes of galaxies in the MaNGA sample, when divided into central and satellite galaxies. 

We find that central galaxies have a wider range in shapes, being less axisymmetric than satellite or isolated galaxies. No clear trend of galaxy shape is observed with halo mass or local density. These results are consistent with \citet{vandenBoschF2008}, who analysed the colours and concentrations of SDSS galaxies and found no difference in the shape of galaxies as a function of halo mass. 


Galaxies with different shapes are predicted to have evolved along different paths. For example, the simulations of \citet{Jesseit2009} and \citet{Moody2014} showed that minor mergers led to flatter remnants (lower q) and higher triaxiality than major mergers. The most triaxial galaxies are usually formed by sequential mergers or re-mergers \citep{Moody2014}. Similarly, \citet{Taranu2013} found that simulations of multiple dry minor mergers usually led to triaxial systems. 
\citet{Lagos2022} used mock observations of galaxies in the EAGLE simulation to predict that flat (low $q$) slow-rotating galaxies are preferentially formed from major mergers; whereas round slow-rotating galaxies are formed from minor or very minor mergers; and prolate slow-rotating galaxies from gas-poor mergers. They also found that flat and prolate galaxies are more common among satellite galaxies in massive haloes (with $M_{200}> 10^{13.6} M_{\odot}$). 

These simulations would suggest that the central galaxies in our sample are consistent with having undergone multiple minor mergers, leading to more triaxial shapes. The satellite galaxies in our sample would have evolved from different paths: slow-rotating, oblate axisymmetric, satellite galaxies could be consistent with having undergone gas-rich major mergers or no mergers and interactions with the central galaxy or the tidal field of the host halo; slow-rotating triaxial/prolate satellite galaxies are consistent with gas-poor minor mergers; and fast-rotating satellite galaxies are consistent with having evolved through a channel dominated by gas accretion, bulge growth and quenching, leading to more oblate and flat shapes. Isolated galaxies, being in general very close to oblate axisymmetric systems, are consistent with a scenario where very few (gas-rich) or no mergers, have affected their evolution.

\subsection{Velocity anisotropy}
We find that the global population of central galaxies is in general radially anisotropic or close to isotropic (with average $\beta_r \sim 0.1$; Fig. \ref{fig:lambda_mass}, panel c), while satellite galaxies have average $\beta_r$ values that are slightly negative ($\sim -0.3$; mildly tangentially anisotropic), and isolated galaxies are tangentially anisotropic at all stellar masses (average $\beta_r \sim -0.5$) . Looking at the correlation between the velocity anisotropy parameter $\beta_r$ and the environmental proxies, we find a weak correlation with local density (Fig. \ref{fig:beta_r_env}), so that, at fixed stellar mass, galaxies (with stellar mass $\log M_{\star}/M_{\odot} < 11$) in higher local densities have greater values of $\beta_r$. However, we note that the most radially anisotropic galaxies are in mid-mass haloes and halo mass (Fig. \ref{fig:beta_r_env}). This might be due to the cluster population being dominated by satellites, while central galaxies (which are generally more radially anisotropic), are only a small fraction of the sample presented here (8\%).

We note that previous studies (e.g. \citealt{Lipka2021}) found biases in the derivation of the velocity anisotropy profiles with Schwarzschild modelling, in particular due to the preference of these models for edge-on inclination views. We find that the derived viewing angles for our sample of SAMI galaxies do not show any significant correlations with stellar mass (Spearman's correlation coefficient $\rho = -0.09$, with a $p$-value of 0.087), or with the velocity anisotropy $\beta_r$ ($\rho = 0.11$, with a $p$-value of 0.006), however, we do see higher uncertainties in the values of $\beta_r$ for galaxies with low derived inclination angles, as expected since lower inclination angles are generally harder to constrain.

\begin{figure*}
\centering
\includegraphics[scale=0.45,trim=0.2cm 0.5cm 0.25cm 0cm , clip=True]{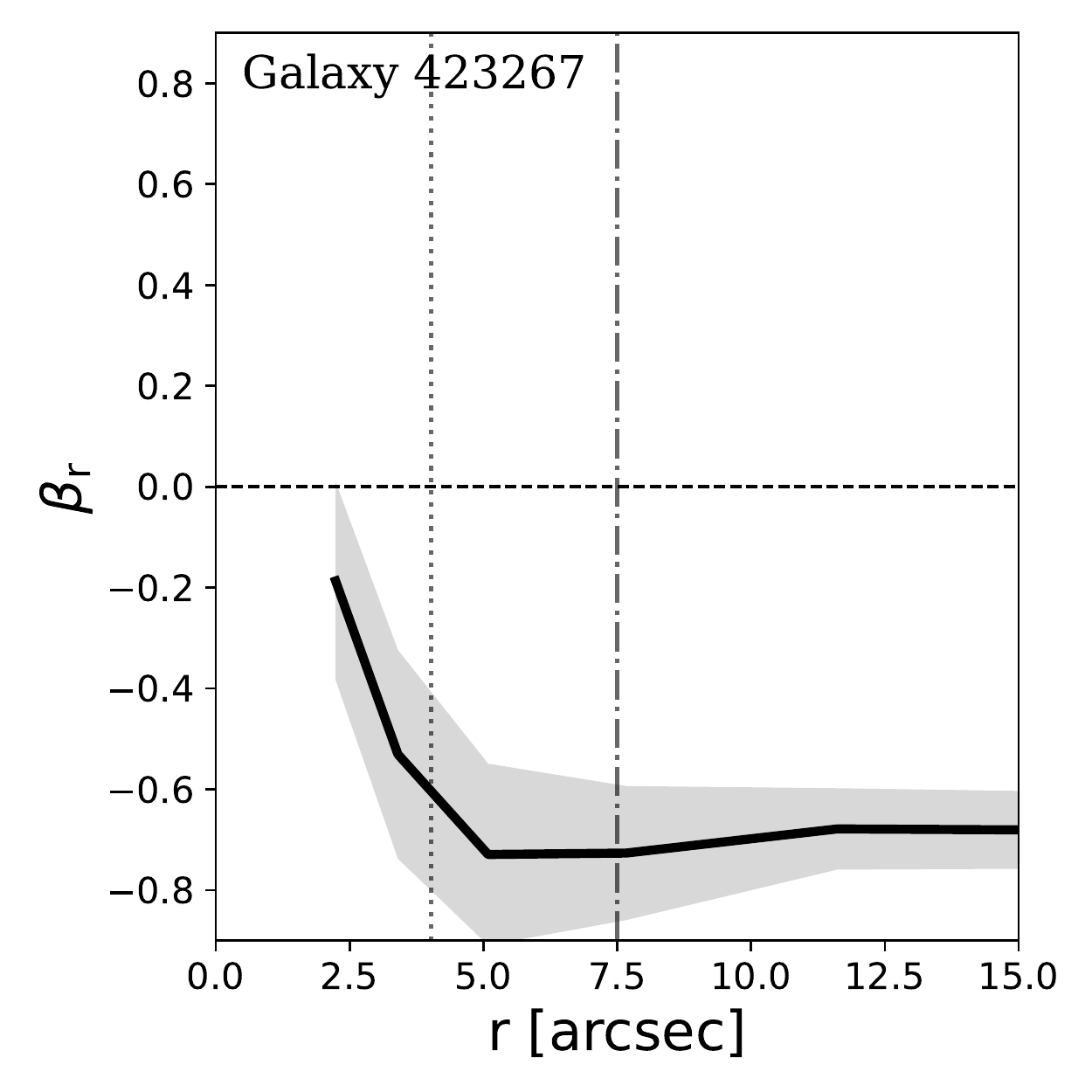}\includegraphics[scale=0.45,trim=0.2cm 0.5cm 0.25cm 0cm , clip=True]{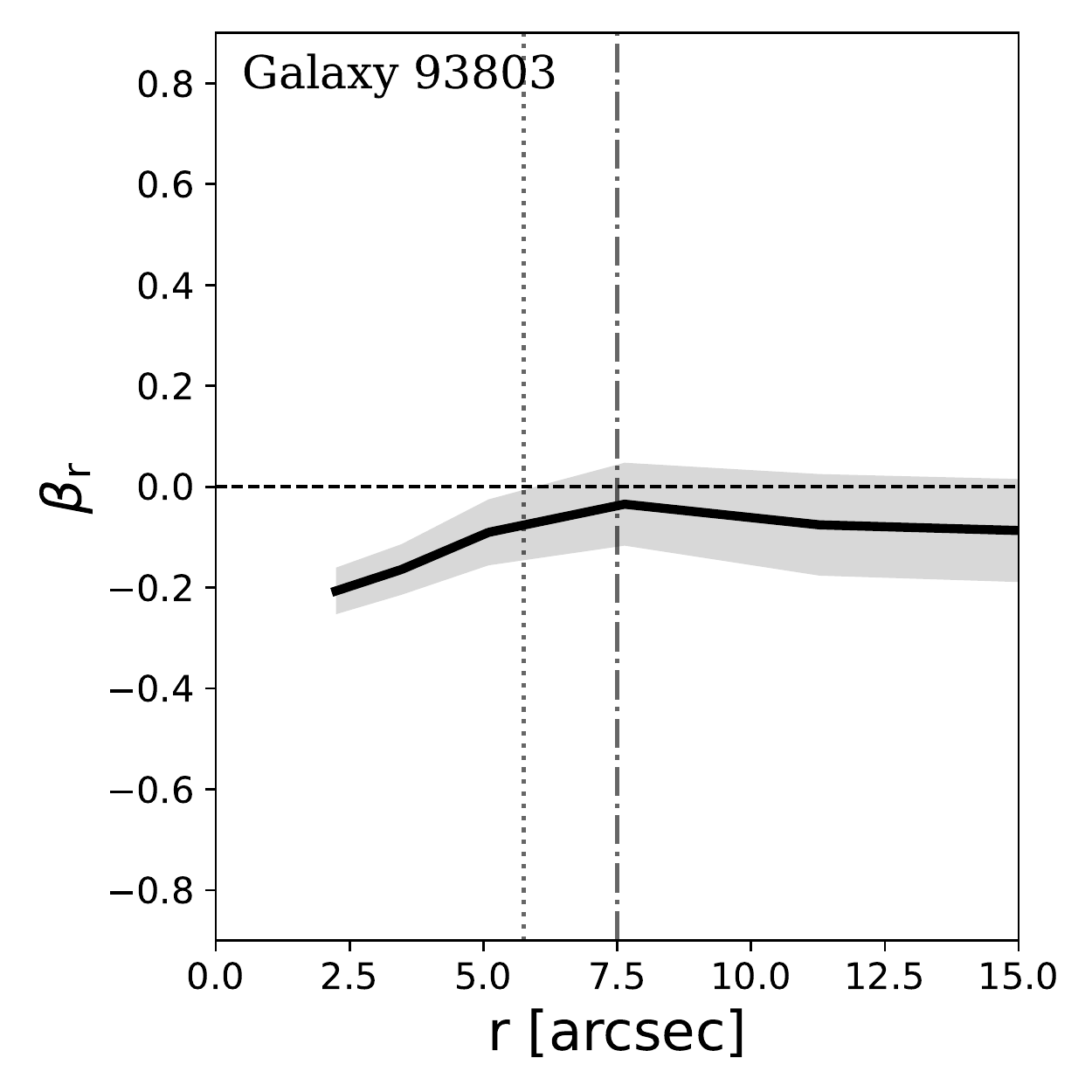} \includegraphics[scale=0.45,trim=0.2cm 0.5cm 0.25cm 0cm , clip=True]{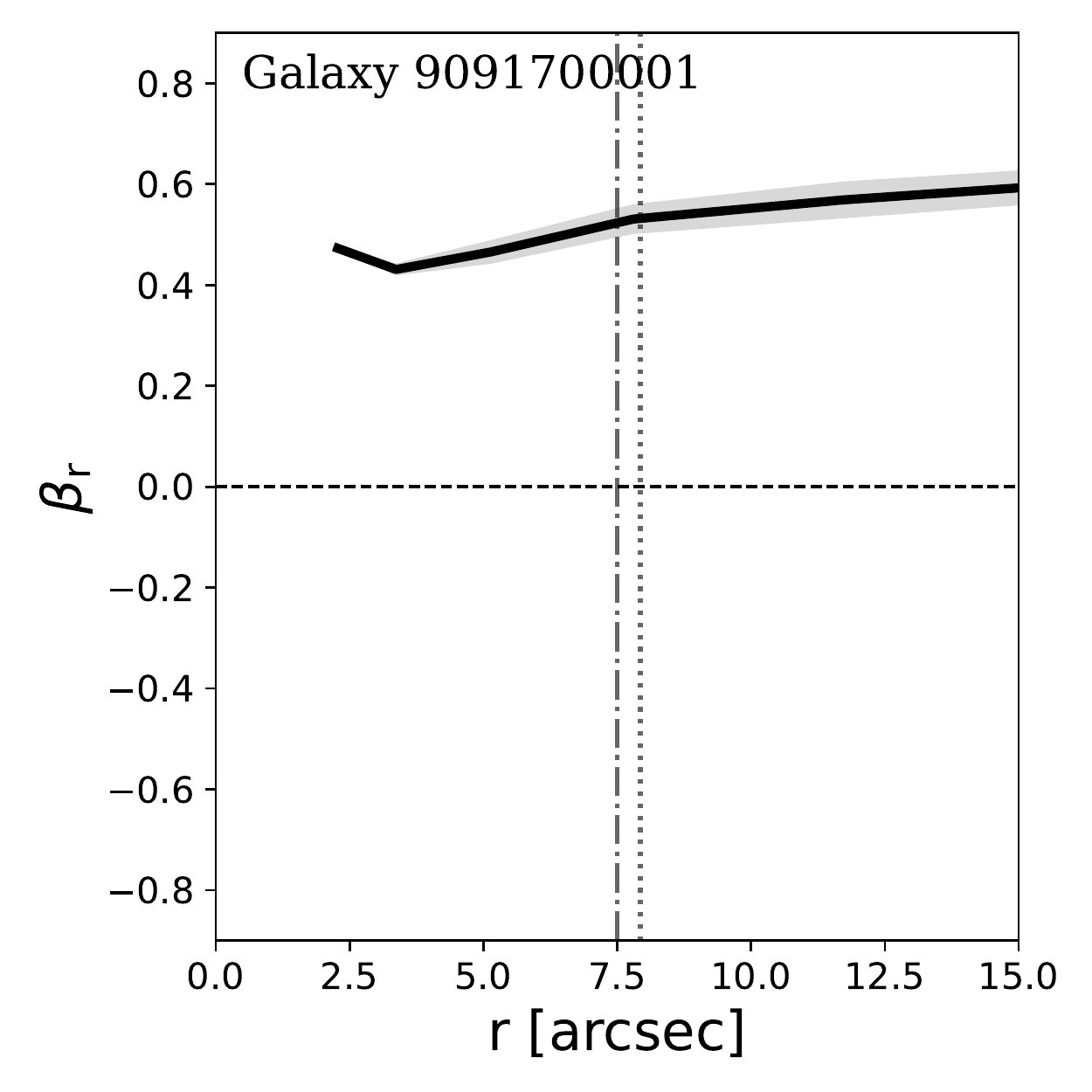}
\caption{Velocity anisotropy profile, $\beta_r$, as a function of the radius for galaxy 423267 (panel a), 93803 (panel b) and 9091700001 (panel c). The solid curves represent the velocity anisotropy profile obtained by the best-fit model. The filled region indicates the errors. The vertical grey dotted and dash-dotted lines are 1$R_{\rm e}$ and $R_{max}$, respectively.}
\label{fig:beta_prof}
\end{figure*}

For completeness, we show in Fig. \ref{fig:beta_prof} the velocity anisotropy profiles for three example galaxies in our sample: one in the low-mass range (panel a - galaxy 423267, $\log M_{\star}/M_{\odot} \sim 10.02$), one in the mid-mass range (panel b - galaxy 93803, $\log M_{\star}/M_{\odot} \sim 10.94$) and one in the high-mass range (panel c - galaxy 909170001, $\log M_{\star}/M_{\odot} \sim 11.41$).

The velocity anisotropy profiles are consistent with those seen in the literature. The $\beta_r$ radial profile of galaxy 9091700001, a high-mass galaxy, has a similar shape to that of NGC 1453, a fast-rotating elliptical galaxy in the MASSIVE Survey \citep{Quenneville2021b}; an increase of $\beta_r$ with the radius, as well as negative radial profiles, were also observed by \cite{Thomas2014}, for a set of high velocity-dispersion elliptical galaxies (without cores). Moreover, similar profiles to ours were also derived by \cite{Jin2019} for nine early-type galaxies from the ILLUSTRIS cosmological simulations \citep{Vogelsberger2014, Genel2014, Nelson2015}.

Simulations show that mergers can have a substantial effect on the anisotropy of the resulting galaxy \citep{Dekel2005}, and a higher fraction of stars accreted from mergers is connected to greater radial anisotropy, while tangential anisotropy is seen only for galaxies with high fractions of stars formed in-situ \citep{Wu2014}. Since galaxies in group environments are predicted to have experienced a greater number of interactions compared to those in less dense environments, an increased rate of mergers would explain the increase in $\beta_r$ that we see with increasing halo mass and local density. 

\citet{Bournaud2007} found, studying galaxies in N-body simulations, that elliptical-like galaxies formed in multiple minor mergers tend to have larger radial anisotropy than those of similar mass formed in one single merger. 
This is consistent with the central galaxies in our sample, which show higher values of velocity anisotropy, having undergone multiple minor mergers during their evolution. The satellite and isolated galaxies, which both show lower values of velocity anisotropy, could be formed with fewer or no minor mergers, or gas-rich major mergers, so that they have greater fractions of stars formed in-situ.


\subsection{Orbital components}
We show the fractions of the warm and hot orbital components, according to their orbit circularity distribution components within 1$R_e$, as a function of stellar mass and environment in Fig. \ref{fig:f_warm_env} and \ref{fig:f_hot_env}. We do not see any significant trend with either halo mass or local density above $\log M_{\star}/M_{\odot} > 11$. However, there is a marginal trend of the fractions of orbits with environment for galaxies with stellar masses below $\log M_{\star}/M_{\odot} \sim 11$, so that galaxies in lower-mass haloes and lower local densities have higher fractions of warm orbits and lower fractions of hot orbits than higher-mass haloes and higher local densities. Central galaxies, at fixed stellar mass, have fewer stars on warm orbits, and large fractions of hot orbits at all stellar masses, but are otherwise consistent with satellite galaxies.

Our results are in general agreement with results from Schwarzschild model fits to early-type galaxies in the MaNGA survey by \citet{Jin2020}. They do not find any significant differences between the orbital components of central and satellite galaxies. When considering the local density environments (indicated by neighbour counts), they found that galaxies that have higher neighbour counts tend to have more hot orbits, consistent with the trend that we see for galaxies in different local densities.

The suggestions we find seem to suggest that, while halo mass and local density might have an effect on lower mass galaxies ($\log M_{\star}/M_{\odot} < 11$), galaxy designation (central/satellite/isolated) is more important for higher-mass galaxies. Similar suggestions of stronger trends with environment for lower-mass galaxies are also found by Turner et al. (in prep), in a sample of 7117 simulated galaxies from EAGLE simulation \citep{Schaye2015} and 3724 observed galaxies from the GAMA survey \citep{Driver2011}. 

The suggestions of orbital fraction trends with halo mass and local density that we see for lower-mass galaxies could be connected with the fraction of accreted stars. The cosmological zoom simulations of \citet{Rottgers2014} found that the accreted star component, having fallen in from all directions, is expected to have more radial orbits, while galaxies with more "late" star-formation formed in-situ due to gas accretion or gas-rich mergers, show a higher fraction of circular orbits. Since the accretion of stars (at low redshift) becomes more important for massive systems \citep[e.g.][]{Guo2008, Oser2010, Naab2014}, we expect massive galaxies to have a higher fraction of stars on hot orbits. \citet{Clauwens2018} studied a sample of galaxies in the EAGLE simulation and found a sharp transition from galaxies dominated by in-situ star-formation to galaxies dominated by ex-situ star-formation at similar masses ($\log M_{\star}/M_{\odot} \geq 11$). This is consistent with empirical results from the GAMA survey \citep{Robotham2014}, which show that the stellar mass assembly in galaxies above $10^{11} M_{\odot}$ is dominated by galaxy mergers. \citet{Clauwens2018} also found that galaxies dominated by in-situ star-formation could be more influenced by non-merger induced tidal perturbations due to their environment, compared to ex-situ dominated galaxies.

The suggestions of weak trends of the orbital fractions with environment for lower-mass galaxies is an interesting indication that needs further study, using a larger sample with a more homogeneous distribution in stellar mass and environment in order to be confirmed.

\subsection{Evolutionary scenarios}
Early-type galaxies show complicated structures in their kinematic maps, but their kinematics can be broadly separated into two classes, fast- and slow-rotating galaxies (for a recent statistical analysis see: \citealt{vandesande2021a}). According to \citet{Cappellari2016}, these two classes also represent two major channels of galaxy formation. In this representations, fast-rotating galaxies start as star-forming disc galaxies and evolve through a set of processes dominated by gas accretion, bulge growth and quenching. By comparison, slow-rotating galaxies assemble near the centre of massive haloes via intense star formation at high redshift, and evolve from a set of processes dominated by gas-poor mergers, resulting in more triaxial shapes. These galaxies are the most massive.

Simulations suggest that there are multiple pathways to creating galaxies in the various kinematic classes \citep[e.g.][]{Naab2014}. For example, one effective way of transforming the kinematics of galaxies is via galaxy mergers \citep[e.g.][]{Jesseit2009, DiMatteo2009, Bois2011, Naab2014, Penoyre2017, Choi2017, Lagos2017, Lagos2018}. For example, \citet{Naab2014} and \citet{Lagos2018a} found that gas poor mergers are an effective way to decrease $\lambda_{Re}$, but also a series of minor mergers or a single major merger can have a similar effect \citep{Naab2014,Choi2017,Schulze2018, Lagos2018a}. 

Our results from \citet{Santucci2022} indicate that the internal structures of galaxies are dominated by the physical processes associated with the growth of stellar mass and are in general agreement with the two formation channels proposed by \citet{Cappellari2016}, where the evolution of massive galaxies is dominated by mergers, leading to the spin-down of the systems. However, we also see intriguing suggestions for a connection between galaxy evolution and environment. We find suggestions of a trend of the orbital fractions with environment, for lower-mass galaxies, so that galaxies in low-mass haloes (or low local densities) are more rotationally-supported (axisymmetric oblate-like galaxies, with tangential anisotropy, higher values of $\lambda_{Re,EO}$, higher fractions of warm orbits and lower fractions of hot orbits) than galaxies in higher-mass haloes (higher local densities). 

Our results suggest two different ways in which environment can affect galaxy evolution. This seems to depend on the galaxy's stellar mass, with a transition at $\log M_{\star}/M_{\odot} \sim 11$. This transition mass is also consistent with the sharp transition from in-situ dominated galaxies to ex-situ dominated galaxies found by \citet{Clauwens2018} and \citet{Cannarozzo2022}, and with the transition mass ($\log M_{\star}/M_{\odot} \sim 11.2$) \citet{Cappellari2016} proposed to mark the increase in the slow-rotating population.

In the scenario suggested by our results, high-mass galaxies (with $\log M_{\star}/M_{\odot} > 11$) evolve from a set of sequential mergers (mostly minor), where the fraction of accreted stars is connected to their internal structure: these galaxies are generally pressure supported, with a high fraction of hot orbits, radial anisotropy and more triaxial shape. In our analysis, these galaxies show differences in structure depending on whether they are central, satellite or isolated galaxies, with central galaxies being the slowest-rotating, radially anisotropic galaxies. This picture is consistent with central galaxies having undergone a high number of mergers, in particular minor mergers, that led to a spin-down and to their present-day structure \citep[e.g.][]{Bournaud2007,Choi2018,Lagos2022}.

In contrast, galaxies with stellar masses below $\log M_{\star}/M_{\odot} \sim 11$ are consistent with being more influenced by non-merger induced tidal perturbations due to their environment. High-speed encounters are more probable in higher-density environments and they can lead to a spin-down of the galaxy and can induce changes in their structures. In our analysis, we see suggestions of these effects in both the velocity anisotropy, the orbital fractions and the edge-on spin-parameter $\lambda_{Re, EO}$, so that at fixed stellar mass, galaxies in low local densities have more tangential velocity anisotropy, higher fractions of warm orbits, lower fractions of hot orbits and are more tangentially supported than galaxies in higher local densities. Suggestions of similar trends are also found with halo mass, but they are less clear, likely because the low-stellar mass / low-halo mass region is not well sampled. 

The suggestions pointing to the environment affecting the structure of lower mass galaxies are very interesting. Exploring these results further needs a larger sample, covering a better-sampled range in both stellar mass and environment. 
We predict the sample size needed in order to observe correlations in these parameters that are significant at the 3-$\sigma$ level. To do this, we create mock-SAMI galaxies by adding random scatter to the galaxy properties found in this analysis, assuming that the suggestions of correlations we observe are true. We find that we need at least 1500 galaxies (a $\sim$10x increase in sample size) in order to find correlations significant at the 3-$\sigma$ level. The upcoming Hector galaxy survey \citep{Bryant2016, Bryant2020}, with over 15000 total galaxies, would be the ideal sample with which to test the intriguing conclusions presented here. Undertaking similar analysis in galaxies at higher redshifts \citep[e.g. using the MAGPI survey - ][]{Foster2021} would also help to isolate the evolutionary effect of environment on the stellar kinematics of galaxies. 

\section{Conclusions}

Using intrinsic galaxy properties derived from building Schwarzschild models, we explore the correlation between the intrinsic shape, velocity anisotropy, orbital components and spin parameter and four environmental proxies: central, satellite or isolated designation, halo mass and local $5^{th}$ nearest neighbour galaxy density. Our sample consists of 153 galaxies from the SAMI Galaxy Survey, with stellar masses ranging from $9.5 < \log (M_{\star}/ M_{\odot}) < 12$. Our key findings are:
\begin{itemize}
    \item Central galaxies (with stellar mass $\log M_{\star}/M_{\odot} > 11$) have lower values of $\lambda_{Re, EO}$ than satellite galaxies of similar stellar mass (Fig. \ref{fig:lambda_mass}, panel a) and galaxies in higher local densities show lower values of \lam, at fixed stellar mass (Fig. \ref{fig:lam_r_env}).
    \item At fixed stellar mass, there is no significant difference in the mean triaxiality of central, satellite and isolated galaxies or for galaxies in different environments (Fig. \ref{fig:lambda_mass}, panel b).
    \item Central galaxies are generally radially anisotropic, while satellite and isolated galaxies are mostly supported by tangential anisotropy (Fig. \ref{fig:lambda_mass}, panel c). The velocity anisotropy parameter, $\beta_r$, shows a weak correlation with environment: in particular with local density (Fig. \ref{fig:beta_r_env}), so that at fixed stellar mass, galaxies in higher local densities have greater values of $\beta_r$.
    \item We do not see any significant trend in the orbital fractions within 1$R_e$ with either halo mass or local density, above $\log M_{\star}/M_{\odot} > 11$. We find interesting suggestions of trends in the fractions of orbits with environment for galaxies with stellar masses below this, so that galaxies in lower-mass haloes (or less dense local environments) have higher fractions of warm orbits and lower fractions of hot orbits than galaxies in higher-mass haloes (or denser local environments), as shown in Figs. \ref{fig:f_warm_env} to \ref{fig:f_hot_env}. 
\end{itemize}

The results presented here support a scenario where environment plays a role in shaping present-day galaxies, and that role is secondary to stellar mass. In particular, we find suggestions consistent with the picture that the evolution of high-mass galaxies ($\log M_{\star}/M_{\odot} > 11$) is merger-driven, with differences in their structure depending on whether they are central, satellite or isolated galaxies. The evolution of lower-stellar mass galaxies ($\log M_{\star}/M_{\odot} < 11$), on the other hand, is consistent with being affected by non-merger tidal perturbations, so that galaxies in denser environments show different structures compared to galaxies in less dense environments.

Due to the size of our sample and the scatter in the presented trends, our results can only provide interesting suggestions that environment is affecting the structure of lower mass galaxies. Larger samples, covering a better-sampled range in both stellar mass and environment are needed to confirm our results.

\section*{Acknowledgements}
We thank the anonymous referee for their comments that helped to improve this manuscript.
GS acknowledges useful discussions with the DYNAMITE team and L. Cortese.
The SAMI Galaxy Survey is based on observations made at the Anglo-Australian Telescope. The Sydney-AAO Multi-object Integral field spectrograph (SAMI) was developed jointly by the University of Sydney and the Australian Astronomical Observatory. The SAMI input catalogue is based on data taken from the Sloan Digital Sky Survey, the GAMA Survey and the VST ATLAS Survey. The SAMI Galaxy Survey is supported by the Australian Research Council center of Excellence for All Sky Astrophysics in 3 Dimensions (ASTRO 3D), through project number CE170100013, the Australian Research Council center of Excellence for All-sky Astrophysics (CAASTRO), through project number CE110001020, and other participating institutions. The SAMI Galaxy Survey website is \href{http://sami-survey.org/}{http://sami-survey.org/}.\\
\\

GS acknowledges support of an Australian Government Research Training Program (RTP) Scholarship and of an Australian Research Council Discovery Project (DP210101945) funded by the Australian Government. 
SB acknowledges funding support from the Australian Research Council through a Future Fellowship (FT140101166).
JvdS acknowledges support of an Australian Research Council Discovery Early Career Research Award (project number DE200100461) funded by the Australian Government.
RMcD acknowledges funding support via an Australian Research Council Future Fellowship (project number FT150100333).
JBH is supported by an ARC Laureate Fellowship FL140100278. The SAMI instrument was funded by Bland-Hawthorn's former Federation Fellowship FF0776384, an ARC LIEF grant LE130100198 (PI Bland-Hawthorn) and funding from the Anglo-Australian Observatory.
JJB acknowledges support of an Australian Research Council Future Fellowship (FT180100231).
M.S.O. acknowledges the funding support from the Australian Research Council through a Future Fellowship (FT140100255).
GvdV acknowledges funding from the European Research Council (ERC) under the European Union's Horizon 2020 research and innovation programme under grant agreement No 724857 (Consolidator Grant ArcheoDyn).
S.K.Y. acknowledges support from the Korean National Research Foundation (NRF-2020R1A2C3003769).

\section*{Data Availability}

All observational data presented in this paper are available from Astronomical Optics’ Data Central service at https://datacentral.org.au/
as part of the SAMI Galaxy Survey Data Release 3. 

Measurements from Schwarzschild models are available
\sendemail{giulia.santucci@uwa.edu.au}{Schw models: data request}{
contacting the corresponding author.}



\bibliographystyle{mnras}
\bibliography{references} 




\appendix
\section{Intrinsic shape} \label{app:T_re}
\begin{figure*}
\centering
\includegraphics[scale=0.6,trim= 2cm 0cm 2cm 1cm , clip=True]{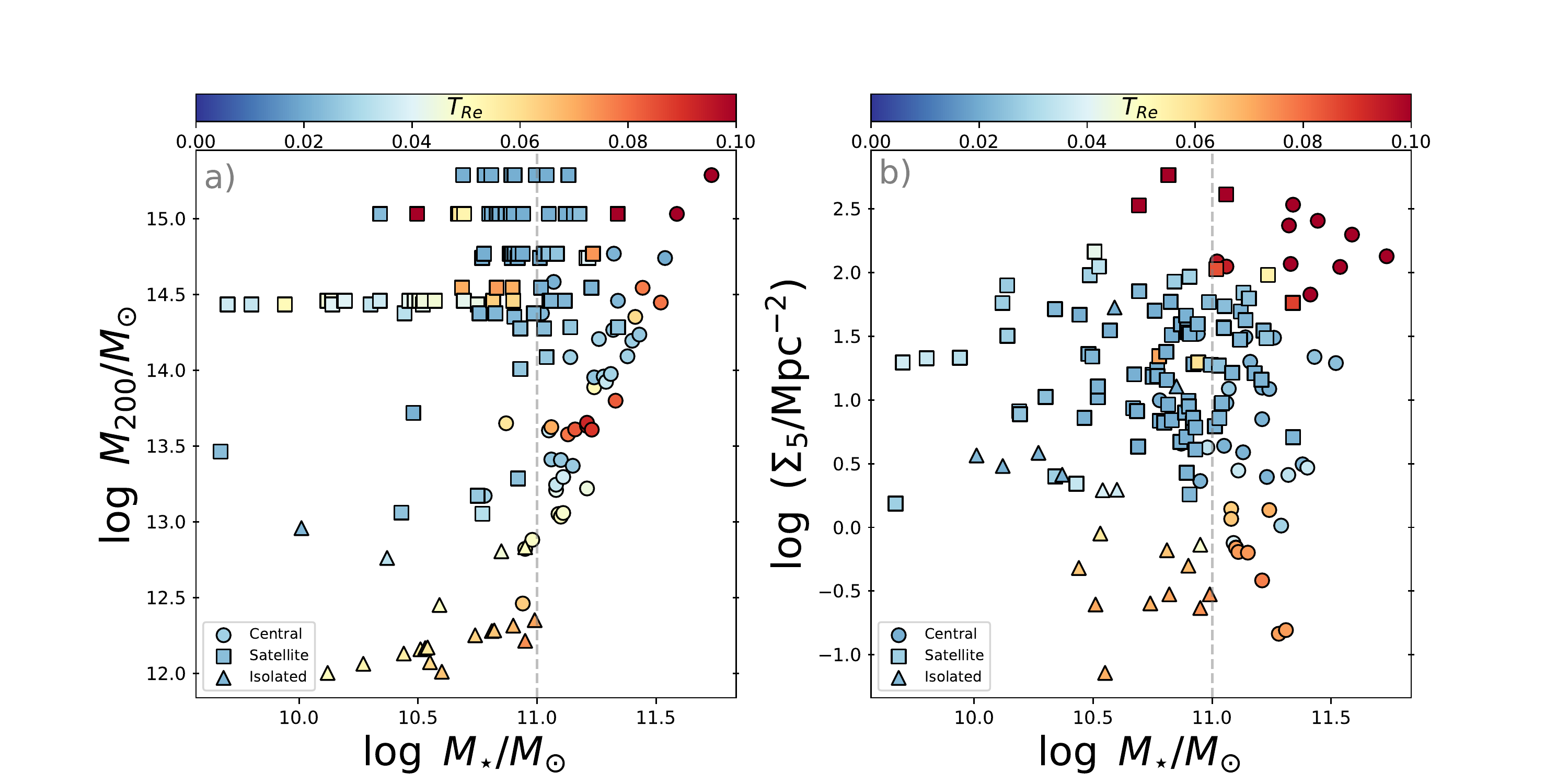} 
\caption{Halo mass $\log \ M_{200}/M_{\odot}$ and local density $\log (\Sigma_5/ \rm Mpc^{-2})$ as a function of stellar mass, colour-coded by triaxiality, LOESS smoothed to reveal any mean underlying trend. Central galaxies are shown as circles, satellite galaxies as squares and isolated galaxies as triangles. We do not find any significant correlation between triaxiality and environment.}
\label{fig:app_triax_env}
\end{figure*}

We show the distribution of galaxy triaxiality in the halo mass-stellar mass and local density-stellar mass plane in Fig. \ref{fig:app_triax_env}. 
In both cases, we do not find any significant correlation between triaxiality and environment.

\section{One-dimensional distributions} \label{app:1dcuts}
\begin{figure*}
\centering
\includegraphics[scale=0.09,trim= 3cm 7cm 1cm 14cm , clip=True]{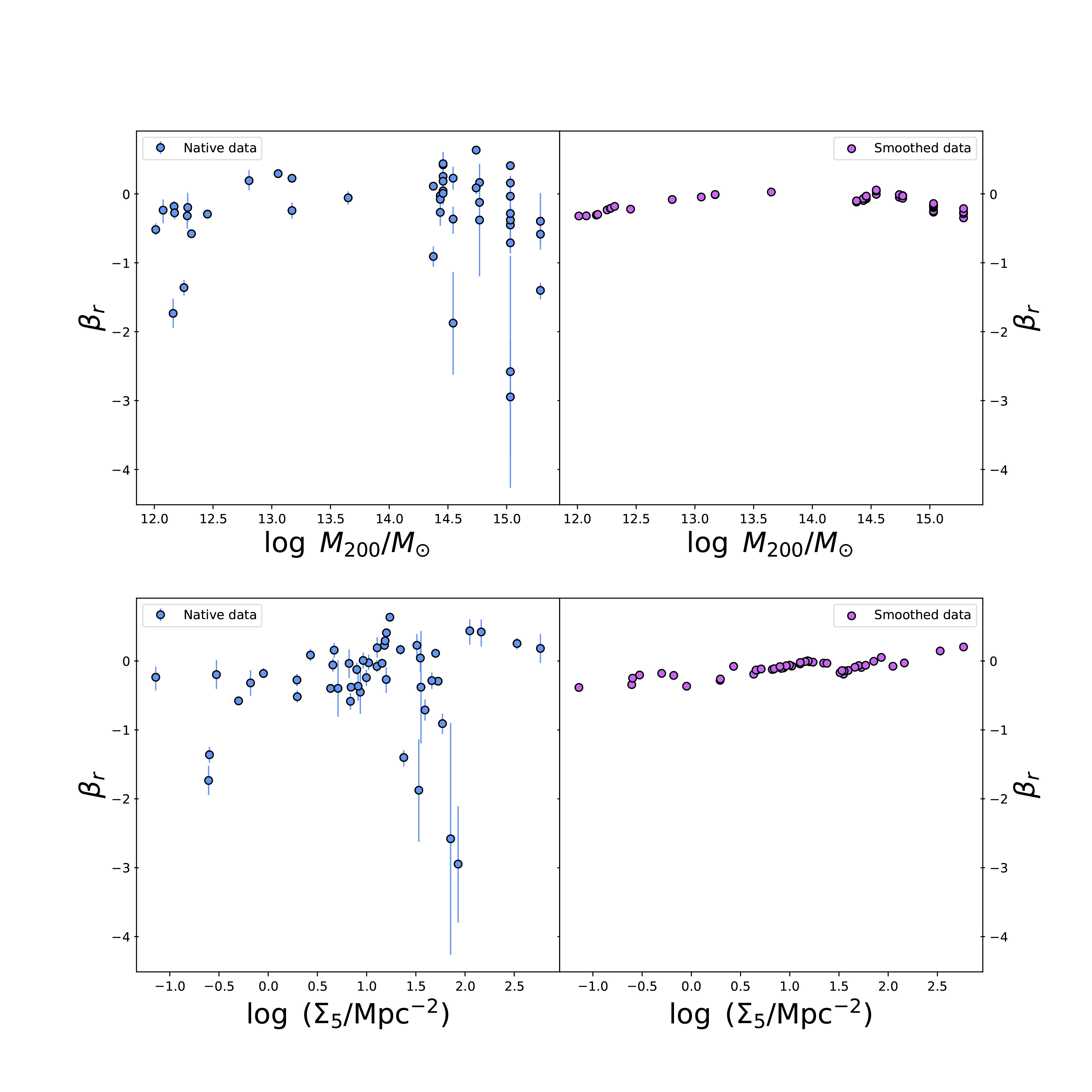}
\caption{The velocity dispersion anisotropy in spherical coordinates $\beta_r$ as a function of halo mass $\log \ M_{200}$ (top rows) and local density $\log (\Sigma_5/ \rm Mpc^{-2})$ (bottom rows). left-hand panels show the native data (in blue0, while the LOESS smoothed data is shown (in purple) in the right-hand panels. The smoothed data is representative, within the uncertainties, of the native data.}
\label{fig:br_1dcut}
\end{figure*}

To better visualise how well the smoothed data reproduce the native data, within the given uncertainties, we show in Fig. \ref{fig:br_1dcut} a one-dimensional cut of $\beta_r$ as a function of halo mass (top row) and as a function of local density (bottom row) for galaxies with stellar mass $\log M_{\star}/M_{\odot} = 10.7 \pm 0.25$. The left-hand panels show the native data (in blue), while the right-hand panels show the smoothed values (in purple) of $\beta_r$.

\begin{figure*}
\centering
\includegraphics[scale=0.09,trim= 3cm 7cm 1cm 15cm , clip=True]{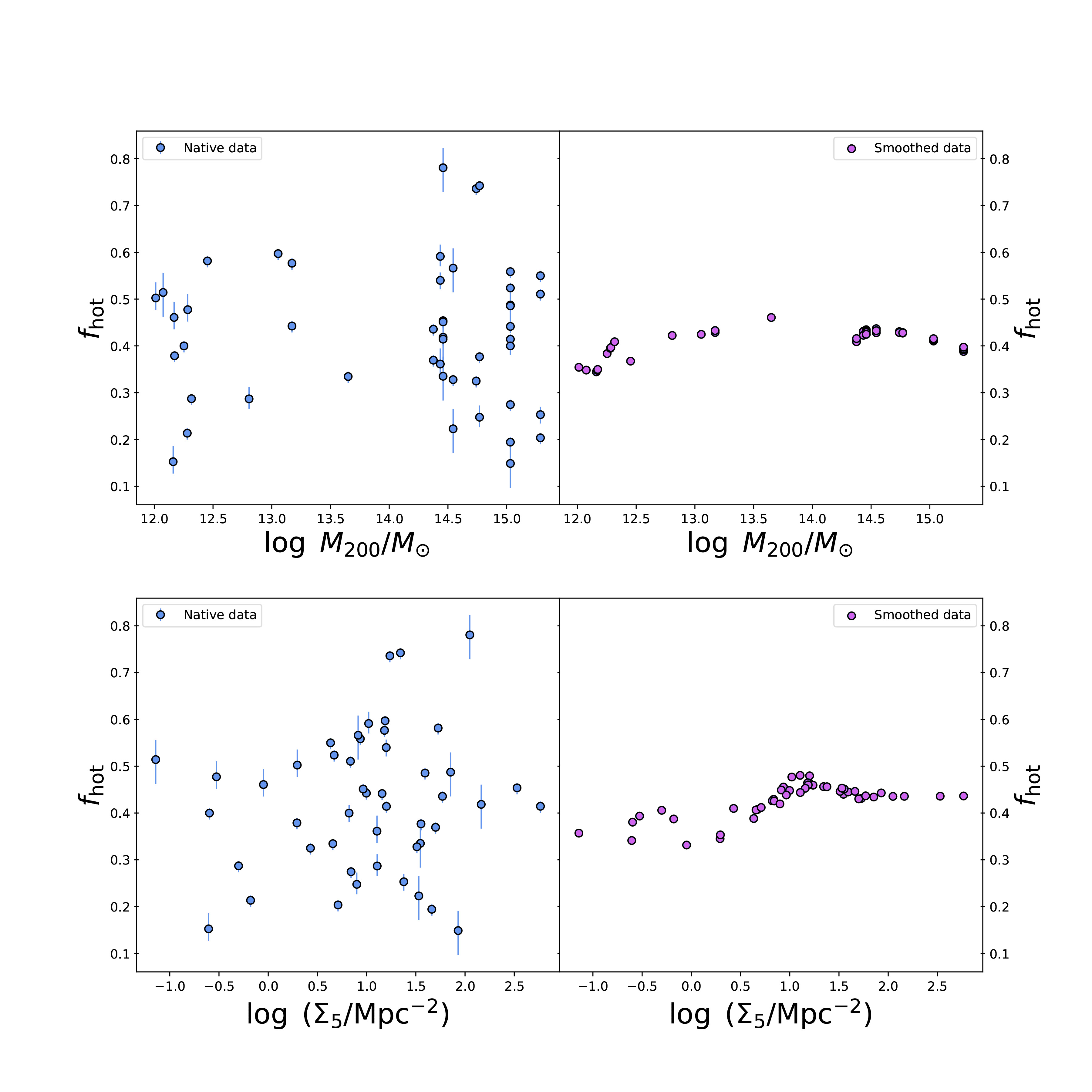}
\caption{Fraction of hot orbits as a function of halo mass $\log \ M_{200}$ (top rows) and local density $\log (\Sigma_5/ \rm Mpc^{-2})$ (bottom rows). left-hand panels show the native data (in blue), while the LOESS smoothed data is shown (in purple) in the right-hand panels. We note that there is significant scatter in the distribution of the fraction of orbits from the native data, compared to the smoothed data.}
\label{fig:fhot_1dcut}
\end{figure*}

\begin{figure*}
\centering
\includegraphics[scale=0.09,trim= 3cm 7cm 1cm 15cm , clip=True]{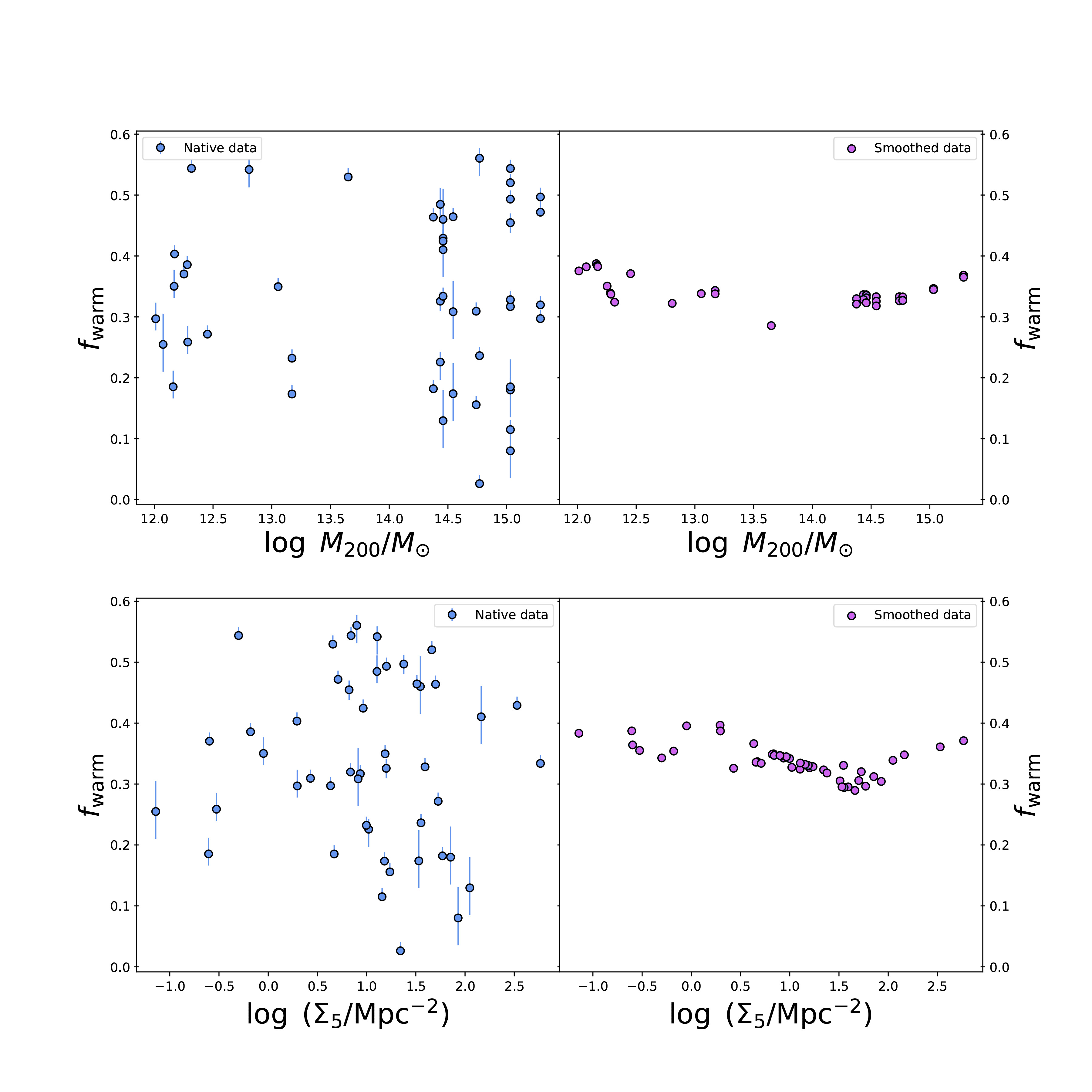}
\caption{Fraction of warm orbits as a function of halo mass $\log \ M_{200}$ (top rows) and local density $\log (\Sigma_5/ \rm Mpc^{-2})$ (bottom rows). left-hand panels show the native data (in blue), while the LOESS smoothed data (in purple) is shown in the right-hand panels. We note that there is significant scatter in the distribution of the fraction of orbits from the native data, compared to the smoothed data.}
\label{fig:fwarm_1dcut}
\end{figure*}

We show in Fig. \ref{fig:fwarm_1dcut} and \ref{fig:fhot_1dcut} a one-dimensional cut of the fraction of warm and hot orbits as a function of halo mass (top row) and as a function of local density (bottom row) for galaxies with stellar mass $\log M_{\star}/M_{\odot} = 10.7 \pm 0.25$. The left-hand panels show the native data (in blue), while the right-hand panels show the smoothed values (in purple) of $\lambda_{Re, EO}$.

\section{Fraction of cold and counter-rotating orbits}\label{app:frac_coldcc}

\begin{figure*}
\centering
\includegraphics[scale=0.6,trim= 2cm 0cm 2cm 1cm , clip=True]{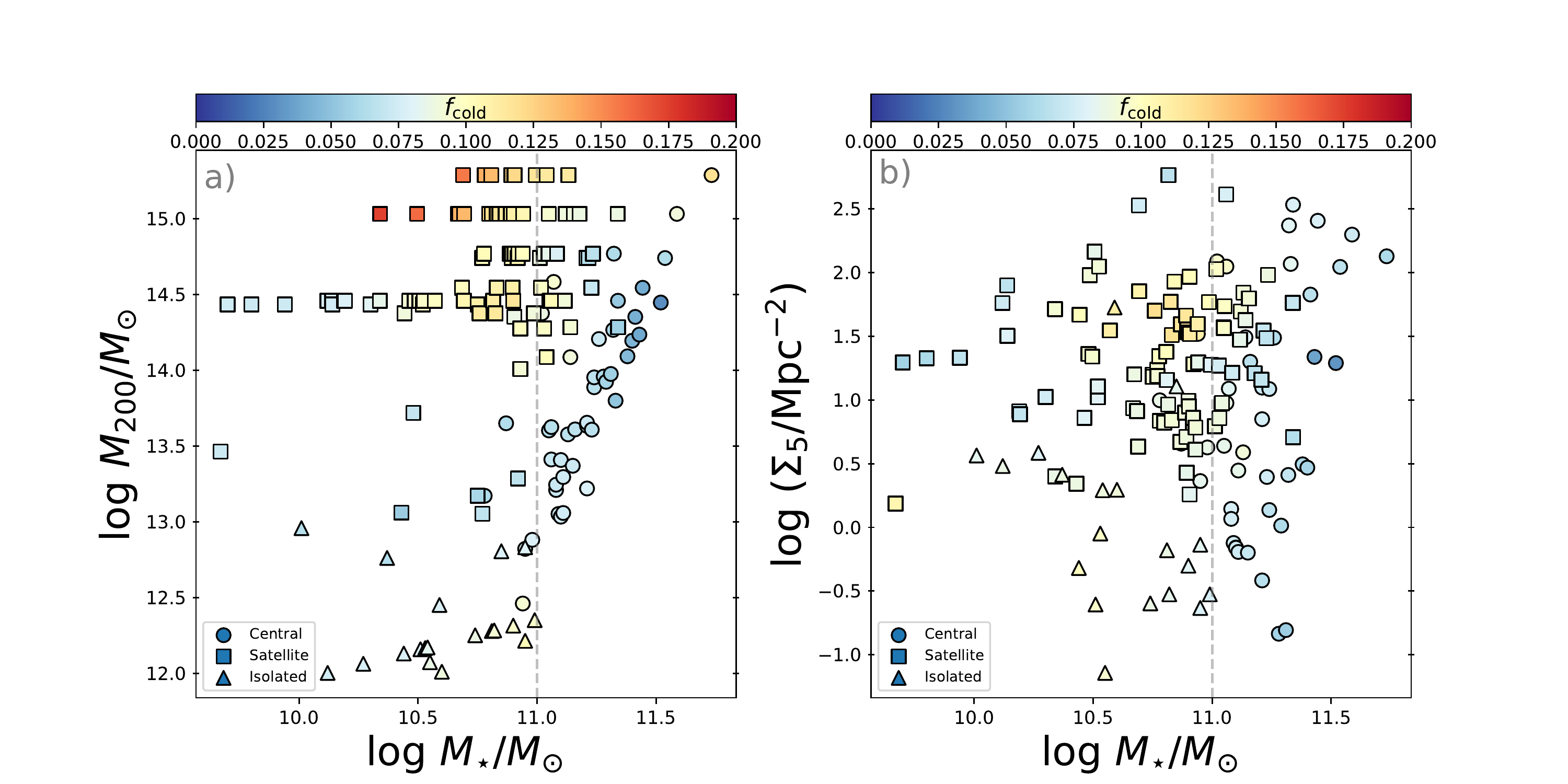} 
\caption{Halo mass $\log \ M_{200}/M_{\odot}$ and local density $\log (\Sigma_5/ \rm Mpc^{-2})$ as a function of stellar mass, colour-coded by the fraction of cold orbits, LOESS smoothed to reveal any mean underlying trend. Central galaxies are shown as circles, satellite galaxies as squares and isolated galaxies as triangles. We do not find any correlation between the fraction of cold orbits and halo mass (panel a) or local density (panel b).}
\label{fig:f_cold_env}
\end{figure*}

When looking at the fraction of cold orbits of galaxies in different halo masses (Fig. \ref{fig:f_cold_env}) we observe that, for stellar masses below $\log M_{\star}/M_{\odot} \sim 11$, galaxies show evidence of a possible additional relationship between the orbital fractions and halo mass so that lower-mass galaxies (below $\log M_{\star}/M_{\odot} \sim 11$) in lower-mass haloes (below $\log M_{200}/M_{\odot} \sim 13.5$) are more likely to have lower fractions of cold orbits. However, we note that the variations in the fractions of cold orbits are small and that our sample does not have many galaxies in the low-stellar mass / low-halo mass region to be able to clearly see a correlation. This suggestion could be connected to satellite galaxies in clusters quenching rapidly and efficiently without undergoing changes in their morphology (e.g. \citealt{Cortese2019}, Turner et al. in prep), therefore retaining their disc-like components. Above $\log M_{\star}/M_{\odot} \sim 11$, any difference in cold orbit fraction between different halo environments disappears. We do not find any additional relationship between the fraction of cold orbits and local density.

\begin{figure*}
\centering
\includegraphics[scale=0.6,trim= 2cm 0cm 2cm 1cm , clip=True]{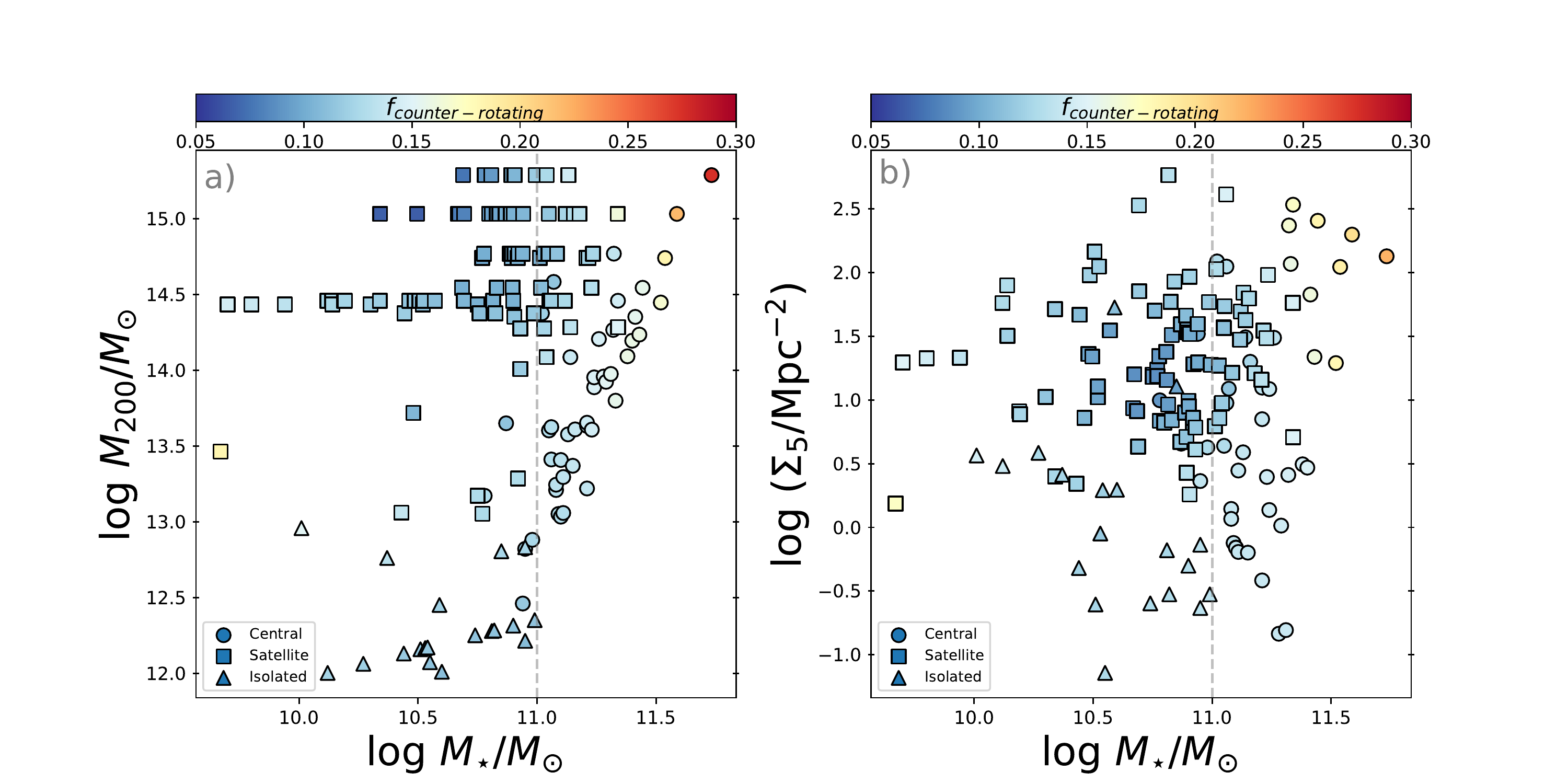} 
\caption{Halo mass $\log \ M_{200}/M_{\odot}$ and local density $\log (\Sigma_5/ \rm Mpc^{-2})$ as a function of stellar mass, colour-coded by the the fraction of counter-rotating orbits, LOESS smoothed to reveal any mean underlying trend. Central galaxies are shown as circles, satellite galaxies as squares and isolated galaxies as triangles. We do not find any correlation between the fraction of counter-rotating orbits and halo mass (panel a) or local density (panel b).}
\label{fig:f_cc_env}
\end{figure*}

We do not find any additional correlation between the fraction of counter-rotating orbits and halo mass or local density (Fig. \ref{fig:f_cc_env}).



\bsp	
\label{lastpage}
\end{document}